\documentclass{aa}
\pdfoutput=1
\usepackage{graphicx}
\usepackage{txfonts}
\usepackage{natbib}
\bibpunct[; ]{(}{)}{;}{a}{}{,}



\begin{document}

   \title{On the Galactic chemical evolution of sulphur}

   \subtitle{Sulphur abundances from the [\ion{S}{i}] $\lambda1082$~nm line in giants
   \thanks{Based partly on observations obtained at the Gemini Observatory, which is operated by the AURA, Inc., under a cooperative agreement with the NSF on behalf of the Gemini partnership: the NSF (US), the PPARC (UK), the NRC (Canada), CONICYT (Chile), the ARC, CNPq (Brazil), and CONICET (Argentina).}\fnmsep
   \thanks{Based partly on observations collected at the European Southern Observatory, Chile (ESO program 080.D-0675).}}

   \author{E. Matrozis
          \inst{1}
          \and
          N. Ryde\inst{1}
          \and
          A. K. Dupree\inst{2}
          }

   \institute{Department of Astronomy and Theoretical Physics, Lund Observatory, Lund University,
              Box 43, 221 00, Lund, Sweden\\
              \email{ryde@astro.lu.se}
        \and
             Harvard-Smithsonian Center for Astrophysics, 60 Garden St., Cambridge, MA 02138, USA\\
             }

   \date{Received ; accepted }

 
  \abstract
   {The Galactic chemical evolution of sulphur is still under debate. At low metallicities some studies find no correlation between [S/Fe] and [Fe/H], which is typical for other $\alpha$-elements, while others find [S/Fe] increasing towards lower metallicities, and still others find a combination of the two. Each scenario has different implications for the Galactic chemical evolution of sulphur.}
   {The aim of this study is to contribute to the discussion on the Galactic chemical evolution of sulphur by deriving sulphur abundances from non-LTE insensitive spectral diagnostics in Disk and Halo stars with homogeneously determined stellar parameters.}
   {We derive effective temperatures from photometric colours, surface gravities from stellar isochrones and Bayesian estimation, and metallicities and sulphur abundances from spectrum synthesis. We derive sulphur abundances from the [\ion{S}{i}] $\lambda1082$~nm line in 39 mostly cool and metal-poor giants, using 1D LTE MARCS model atmospheres to model our high-resolution near-infrared spectra obtained with the VLT, NOT and Gemini South telescopes.}
   {We derive homogeneous stellar parameters for 29 of the 39 stars. Our results argue for a chemical evolution of sulphur that is typical for $\alpha$-elements, contrary to some previous studies that have found high sulphur abundances ($\mathrm{[S/Fe]}\gtrsim0.6$) for stars with $-2.5<\mathrm{[Fe/H]}<-1$. Our abundances are systematically higher by about 0.1~dex in comparison to other studies that arrived at similar conclusions using other sulphur diagnostics.}
   {We find the [\ion{S}{i}] line to be a valuable diagnostic of sulphur abundances in cool giants down to $\mathrm{[Fe/H]}\simeq-2.3$. We argue that a homogeneous determination of stellar parameters is necessary, since the derived abundances are sensitive to them. Our results ($\mathrm{[S/Fe]}$) show reasonable agreement with predictions of contemporary models of Galactic chemical evolution. In these models sulphur is predominantly created in massive stars by oxygen burning, and ejected in the interstellar medium during Type~II supernovae explosions. Systematic differences with previous studies likely fall within modelling uncertainties.}


   \maketitle
%

\section{Introduction}

Sulphur is a chemical element of considerable scientific interest. First, it is one of the $\alpha$-elements, which are elements from O to Ti with even atomic numbers. Studies of $\alpha$-element abundances in stellar atmospheres can shed light on such important properties of stellar populations as their star-formation history and initial mass function \citep[e.g.,][]{Mcwilliam97}. These properties are needed for the discussion on how galaxies are formed and how they evolve.

   Second, sulphur is a volatile element. As such, it does not form dust easily and the number of its atoms measured in a gas is indeed the total number, which makes sulphur suitable for cosmological studies. In particular, it (together with another volatile, zinc) could be used as a cosmological clock in tracing the evolution of damped Ly$\alpha$ systems \citep{Nissen04,Nissen07} -- huge clouds of predominantly neutral hydrogen gas at high redshifts that are thought to play an instrumental role in the formation of galaxies. Before this discussion can be had, however, the chemical evolution of sulphur has to be understood in our own Galaxy.

   Observational studies of Galactic chemical evolution of sulphur were scarce before early 2000s due to the lack of suitable spectral diagnostics, and concerned mainly stars with $\mathrm{[Fe/H]}>-1$ \citep[e.g.,][]{Wallerstein64,Francois87,Francois88}. This has changed partly due to the advent of infrared detector arrays. Now several studies on sulphur abundances in metal-poor stars have been published and left the matter of whether sulphur is a typical $\alpha$-element controversial. To summarize (see \citet{Jonsson11} for a more in-depth review of the methods and findings of previous studies), early studies by \citet{Israelian01} and \citet{TakH02} found that the relative sulphur abundance continuously increases in the metallicity range from solar down to at least $\mathrm{[Fe/H]}\simeq-2.5$, implying a different chemical evolution history from that of other $\alpha$-elements which display a (nearly) metallicity-independent enrichment with respect to iron in stars of low ($\mathrm{[Fe/H]}\lesssim-1$) metallicity \citep{Cayrel04,Jonsell05}.\footnote{The abundance of an element A with respect to element B is $\mathrm{[A/B]}=\log_{10}\left(N_\mathrm{A}/N_\mathrm{B}\right)_\star-\log_{10}\left(N_\mathrm{A}/N_\mathrm{B}\right)_\sun$ where $N$ is the number density. We call [Fe/H] the metallicity and [S/Fe] the sulphur abundance.} A high rate of sulphur-rich hypernovae \citep[e.g.,][]{Nakamura01} and a time-delayed deposition of iron into the interstellar medium \citep{Ramaty00} have been proposed as possible explanations for these unexpected results. On the other hand, a number of studies \citep[e.g,][]{Ryde04,Korn05,Nissen07,Spite11} have found little to no evidence of high ($\mathrm{[S/Fe]}\gtrsim0.6$) sulphur abundances at any metallicity, instead arguing for sulphur being a fairly typical $\alpha$-element and forming a plateau of $\mathrm{[S/Fe]}\sim0.3$ below $\mathrm{[Fe/H]}\sim-1$, which points to an origin in Type~II supernovae \citep[e.g.,][]{Kobayashi06}. Finally, a combination of the two patterns, which implies a very complex history of chemical evolution of sulphur, has been found by \citet{Caffau05} and \citet{Caffau10}.   
   
   The question of whether there exist metal-poor stars with extremely high sulphur abundances is open to debate, and the discrepant findings of previous studies could either reflect reality or differences in the employed analyses. Among these are the differences in sulphur diagnostics employed, stellar parameters adopted and the types of stars analysed, and assumptions made regarding the dimensionality and local thermodynamic equilibrium \citep[LTE; for a discussion on the effects of these assumptions see, e.g.,][]{TakH05,Takeda05,Jonsson11}. In this 1D LTE study of sulphur abundances in cool giants we aim to mitigate some of these concerns:
\begin{enumerate}
\item instead of adopting the stellar parameters ($T_\mathrm{eff}$, $\log{g}$ and $\mathrm{[Fe/H]}$) from multiple previous studies, as most of the studies concerning sulphur have done, we determine them using a single methodology to avoid introducing additional scatter in the results;
\item only the [\ion{S}{i}] $\lambda1082$~nm line is used to derive sulphur abundances. This forbidden line is insensitive to the assumption of LTE putting at least one of the two main modelling assumptions -- radial symmetry and local thermodynamic equilibrium -- on a more solid footing.
\end{enumerate}

\section{Observations}\label{ch:obs}

   While we are mainly interested in sulphur abundances in stars with $\mathrm{[Fe/H]}<-1$, about half of our sample consists of more metal-rich stars. The sample is combined from four subsamples: 1)~14 ``disk'' ($\mathrm{[Fe/H]}>-1$) stars from \citet{Ryde06} observed with the Gemini South telescope (Phoenix spectrometer) with $R=\lambda/\Delta\lambda\simeq60\,000$ and $\mathrm{S/N}\simeq200$; 2)~the metal-poor ($\mathrm{[Fe/H]}<-1$) sample observed by \citeauthor{Ryde06} within the same programme, but previously unpublished; 3)~data from \citet{Jonsson11} obtained at VLT (CRIRES) with $R\simeq80\,000$ and S/N between about 250 and 500 for most stars (HD~13979 and HD~103545 have S/N of 570 and 160, respectively); 4)~three metal-deficient ($\mathrm{[Fe/H]}\sim-0.4$) stars observed by \citet{Ilyin00} with the NOT (SOFIN) with $R\simeq80\,000$ and $\mathrm{S/N}\simeq70$. The wavelength coverage is about 5~nm ($\lambda\simeq1080$--$1085$~nm) for the Phoenix and SOFIN data and about 10~nm ($\lambda\simeq1080$--$1092$~nm with a 2~nm gap around 1086~nm) for the CRIRES data. We refer to the publications of \citet{Ryde06}, \citet{Jonsson11} and \citet{Ilyin00} for detailed descriptions of subsamples 1-2, 3 and 4, respectively. Basic data of the sample stars is given in Table~\ref{tbl:sample}.

\begin{table*}
\caption{\label{tbl:sample}Basic data of the analysed stars.}
\centering
\begin{tabular}{r l c c r @{.} l c c c c c c}
\hline\hline
\multicolumn{2}{c}{\,\,\,Star identifier} & RA(J2000)$^a$ & Dec(J2000)$^a$ & \multicolumn{2}{c}{$\pi^a$} & $V^b$ & $B-V^b$ & Ref.$^b$ & $E_{B-V}^c$ & $JHK_\mathrm{s}$ & Subsample$^e$ \\
HD/BD & HIP & (h m s) & (d m s) & \multicolumn{2}{c}{(mas)} & & &  & & flag$^d$ &\\
\hline
2796    & 2463   & 00 31 16.915 & $-16$ 47 40.79 & $ 0$ & $88\pm0.81$  & 8.495 & 0.747 &  1  & 0.020 & AAA & 2  \\
3546    & 3031   & 00 38 33.345 & $+29$ 18 42.28 & $19$ & $91\pm0.19$  & 4.361 & 0.871 &  1  & 0.008 & TCS & 1  \\
8724    & 6710   & 01 26 17.594 & $+17$ 07 35.11 & $ 2$ & $52\pm0.71$  & 8.300 & 0.987 &  1  & 0.106 & AAA & 2  \\
10380   & 7884   & 01 41 25.894 & $+05$ 29 15.39 & $ 8$ & $98\pm0.23$  & 4.437 & 1.364 &  1  & 0.016 & TCS & 1  \\
13979   & 10497  & 02 15 20.853 & $-25$ 54 54.86 & $ 0$ & $46\pm1.13$  &\ldots &\ldots &\ldots & 0.012 & AAA & 3  \\
21581   & 16214  & 03 28 54.486 & $-00$ 25 03.09 & $ 4$ & $03\pm1.00$  & 8.710 & 0.825 &  1  & 0.069 & AAA & 3  \\
23798   & 17639  & 03 46 45.722 & $-30$ 51 13.35 & $ 1$ & $83\pm0.74$  & 8.302 & 1.098 &  1  & 0.008 & AAA & 3  \\
26297   & 19378  & 04 09 03.418 & $-15$ 53 27.07 & $ 1$ & $59\pm0.78$  & 7.470 & 1.110 &  1  & 0.030 & AEA & 3  \\
29574   & 21648  & 04 38 55.733 & $-13$ 20 48.13 & $ 1$ & $40\pm0.88$  & 8.336 & 1.401 &  1  & 0.183 & AEA & 3  \\
34334   & 24727  & 05 18 10.569 & $+33$ 22 17.86 & $14$ & $04\pm0.58$  & 4.538 & 1.266 &  1  & 0.014 & TCS & 4  \\
36702   & 25916  & 05 31 52.230 & $-38$ 33 24.04 & $ 0$ & $46\pm0.58$  & 8.365 & 1.216 &  1  & 0.028 & AAA & 3  \\
37160   & 26366  & 05 36 54.389 & $+09$ 17 26.41 & $27$ & $76\pm0.27$  & 4.082 & 0.958 &  1  & 0.008 & TCS & 4  \\
40460   & 28417  & 06 00 06.039 & $+27$ 16 19.87 & $ 6$ & $90\pm0.56$  & 6.609 & 1.022 &  1  & 0.026 & CCD & 1  \\
44007   & 29992  & 06 18 48.528 & $-14$ 50 43.44 & $ 5$ & $57\pm0.84$  & 8.059 & 0.839 &  1  & 0.033 & AAA & 3  \\
65953   & 39211  & 08 01 13.336 & $-01$ 23 33.37 & $ 7$ & $25\pm0.30$  & 4.676 & 1.491 &  1  & 0.009 & DDD & 4  \\
81192   & 46155  & 09 24 45.336 & $+19$ 47 11.88 & $ 8$ & $62\pm0.46$  & 6.530 & 0.947 &  1  & 0.016 & TCS & 1  \\
83212   & 47139  & 09 36 19.952 & $-20$ 53 14.74 & $ 0$ & $96\pm0.77$  & 8.335 & 1.070 &  1  & 0.059 & AAA & 3  \\
85773   & 48516  & 09 53 39.242 & $-22$ 50 08.41 & $ 3$ & $41\pm1.20$  & 9.380 & 1.120 &  1  & 0.027 & AAA & 3  \\
103545  & 58139  & 11 55 27.163 & $+09$ 07 45.00 & $ 0$ & $12\pm1.35$  & 9.200 & 0.710 &  2  & 0.034 & AAA & 3  \\
110184  & 61824  & 12 40 14.079 & $+08$ 31 38.07 & $ 0$ & $76\pm0.84$  & 8.305 & 1.175 &  1  & 0.020 & AAA & 2  \\
111721  & 62747  & 12 51 25.196 & $-13$ 29 28.22 & $ 4$ & $33\pm0.86$  & 7.971 & 0.799 &  1  & 0.038 & AAA & 2  \\
117876  & 66086  & 13 32 48.214 & $+24$ 20 48.31 & $ 7$ & $43\pm0.57$  & 6.092 & 0.965 &  1  & 0.008 & TCS & 1  \\
122563  & 68594  & 14 02 31.846 & $+09$ 41 09.95 & $ 4$ & $22\pm0.35$  & 6.200 & 0.904 &  1  & 0.019 & TCS & 2  \\
122956  & 68807  & 14 05 13.026 & $-14$ 51 25.46 & $ 3$ & $16\pm0.59$  & 7.220 & 1.010 &  1  & 0.065 & AEA & 2  \\
139195  & 76425  & 15 36 29.579 & $+10$ 00 36.62 & $14$ & $11\pm0.30$  & 5.260 & 0.945 &  1  & 0.011 & TCS & 1  \\
161074  & 86667  & 17 42 28.362 & $+24$ 33 50.58 & $ 8$ & $04\pm0.36$  & 5.525 & 1.452 &  1  & 0.028 & DCD & 1  \\
166161  & 88977  & 18 09 40.687 & $-08$ 46 45.60 & $ 4$ & $56\pm0.84$  & 8.120 & 0.976 &  1  & 0.293 & AAA & 2  \\
168723  & 89962  & 18 21 18.601 & $-02$ 53 55.68 & $53$ & $93\pm0.18$  & 3.254 & 0.941 &  1  & 0.016 & TCS & 1  \\
184406  & 96229  & 19 34 05.353 & $+07$ 22 44.21 & $30$ & $31\pm0.24$  & 4.450 & 1.177 &  1  & 0.005 & TCS & 1  \\
187111  & 97468  & 19 48 39.575 & $-12$ 07 19.74 & $ 1$ & $45\pm0.82$  & 7.720 & 1.225 &  1  & 0.124 & AEA & 2  \\
188512  & 98036  & 19 55 18.792 & $+06$ 24 24.45 & $73$ & $00\pm0.20$  & 3.715 & 0.855 &  1  & 0.003 & DCD & 1  \\
204543  & 106095 & 21 29 28.213 & $-03$ 30 55.38 & $-0$ & $13\pm1.08$  & 8.600 & 0.760 &  2  & 0.034 & AAA & 2  \\
212943  & 110882 & 22 27 51.522 & $+04$ 41 44.41 & $21$ & $99\pm0.37$  & 4.794 & 1.051 &  1  & 0.021 & TCS & 1  \\
214567  & 111810 & 22 38 52.589 & $+19$ 31 20.16 & $ 8$ & $74\pm0.44$  & 5.820 & 0.925 &  1  & 0.013 & DCD & 1  \\
216143  & 112796 & 22 50 31.089 & $-06$ 54 49.54 & $ 0$ & $87\pm0.91$  & 8.200 & 0.940 &  2  & 0.039 & AAA & 2  \\
219615  & 114971 & 23 17 09.938 & $+03$ 16 56.23 & $23$ & $64\pm0.18$  & 3.696 & 0.921 &  1  & 0.015 & DDD & 1  \\
220954  & 115830 & 23 27 58.096 & $+06$ 22 44.37 & $21$ & $96\pm0.25$  & 4.279 & 1.076 &  1  & 0.014 & TCS & 1  \\
221170  & 115949 & 23 29 28.809 & $+30$ 25 57.86 & $ 2$ & $94\pm0.69$  & 7.674 & 1.085 &  1  & 0.086 & AAA & 2  \\
+30~2611 & 73960 & 15 06 53.830 & $+30$ 00 36.95 & $ 1$ & $07\pm1.23$ & 9.123 & 1.240 &  1  & 0.021 & AAA & 2  \\
\hline
\end{tabular}
\tablefoot{$^a$ taken from \citet{vanLeeuwen07}; $^b$ taken from: 1 -- \citet{GCPD97}, 2 -- \citet{Rossi05}; $^c$ derived in this work (see Sect.~\ref{ch:meth_spar_teff}); $^d$ 2MASS photometry quality flags for colours $J$, $H$ and $K_\mathrm{s}$ \citep{Skrutskie06}. TCS -- colours transformed to 2MASS system from the TCS system (see text). The $H$ magnitude of HD~26297, 29574 and 187111 was also transformed from the TCS system; $^e$ as defined in Sect.~\ref{ch:obs}.}
\end{table*}

\section{Analysis}

In this section we describe the determination of stellar parameters (Sect.~\ref{ch:meth_spar}) and sulphur abundances (Sect.~\ref{ch:meth_sfe}).

\subsection{Stellar parameters}\label{ch:meth_spar}

\subsubsection{Effective temperature}\label{ch:meth_spar_teff}
    The effective temperatures were derived from the photometric colour-effective temperature calibrations of \citet{GHB09}. In particular, we used photometric data from the Johnson and 2MASS photometric systems to derive $T_\mathrm{eff}$ from the $B-V$, $V-J$, $V-H$, $V-K_\mathrm{s}$ and $J-K_\mathrm{s}$ colours.
    
    Each colour gives an estimate of effective temperature $T_{\mathrm{eff},i}=5040~\mathrm{K}/\theta_{\mathrm{eff},i}$, where
\begin{equation}\label{eqn:teff}
\theta_{\mathrm{eff},i}=a_0+a_1X_i+a_2X_i^2+a_3X_i\mathrm{[Fe/H]}+a_4\mathrm{[Fe/H]}+a_5\mathrm{[Fe/H]}^2,
\end{equation}
$X_i$ is the (de-reddened) colour, and $a_j$ are the coefficients of the respective calibration. We then adopted a weighted mean effective temperature according to
\begin{equation}\label{eqn:teffmean}
\bar{T}_\mathrm{eff}=\frac{\sum w_iT_{\mathrm{eff},i}}{\sum{w_i}},
\end{equation}
where $w_i=(\Delta T_{\mathrm{eff},i})^{-2}$. The uncertainties in the individual estimates ($\Delta T_{\mathrm{eff},i}$) are formally propagated uncertainties in the colours, metallicities, colour excesses, and the calibrations themselves. We adopted the weighted mean of the squared deviations from $\bar{T}_\mathrm{eff}$ as the uncertainty of the temperatures:
\begin{equation}\label{eqn:T_meanerr}
\Delta\bar{T}_\text{eff}=\frac{\sum w_i(T_{\mathrm{eff},i}-\bar{T}_\mathrm{eff})^2}{\sum{w_i}}.
\end{equation}

    The Johnson $V$ magnitudes and $B-V$ colours and their uncertainties were taken from the General Catalogue of Photometric Data \citep[GCPD;][]{GCPD97}. The mean uncertainty in $V$ and $B-V$ for the sample is $\sim0.014$ and $\sim0.008$\,mag, respectively, which motivated our assumption of $\Delta V=0.015$ and $\Delta(B-V)=0.010$\,mag for stars that had no error estimates in the catalogue. A few stars (see Table~\ref{tbl:sample}) had no entries in GCPD, and their $V$ magnitudes and $B-V$ colours were taken from \citet{Rossi05}.
    
    The 2MASS magnitudes and their uncertainties were taken from the 2MASS All-Sky Point Source Catalog \citep{Skrutskie06}. The 2MASS magnitudes generally have good precision (quality flag A) for $JHK_\mathrm{s}\ga 5$, which is the case for about a half of our sample. For the other half the uncertainties are considerably larger, introducing an uncertainty of a few hundred K in each $T_{\mathrm{eff},i}$. For these stars we chose to invert the colour transformations of \citet{RM05a} to derive the 2MASS $JHK_\mathrm{s}$ magnitudes from the $JHK$ magnitudes of the TCS photometric system \citep{Alonso94,Alonso98}. The agreement between measured and calculated 2MASS magnitudes for faint stars was found to be excellent. While there is no good reason to assume that the transformation is valid for bright stars, we find good agreement between measured 2MASS (that have large uncertainties) and calculated-from-TCS 2MASS (small uncertainties) magnitudes for them, which indicates that the extrapolation is at the very least valid within the uncertainties of the observed magnitudes. Hence, in order of preference we adopted: 1)\,measured 2MASS magnitudes (when the quality flag was A); 2)\,TCS magnitudes transformed to the 2MASS system (when the quality flag was C or worse); 3)\,2MASS magnitudes with quality flags of C or worse when no TCS data were available.
    
    The colours were de-reddened using the 3D Galactic extinction model of \citet{Drimmel03}. We used the IDL routine provided by the authors of the model to find $A_V$ along the line of sight to the position of the star,\footnote{ftp://ftp.to.astro.it/astrometria/extinction/} assumed $R_V=3.1$ \citep{Cardelli89} to calculate $E_{B-V}=A_V/R_V$, and adopted the extinction ratios $k=E_X/E_{B-V}$ from \citet{RM05b} to calculate the colour excess for the other colours.

    The uncertainties of $E_X$ are large and difficult to estimate due to $A_V$, $\pi$, $R_V$ and $k$ all having substantial uncertainties. \citet{Drimmel03} state that $\Delta A_V$ due to the model can be as large as $0.2A_V$. We estimated the influence of the parallax uncertainties by calculating $A_V$ at the distance extremes allowed by the parallax uncertainties and found additional relative uncertainty of $\Delta A_V/A_V\simeq20$\% on average. Finally, while the assumption of a constant $R_V$ is not well justified, it is unlikely to deviate from the assumed value by more than 10\% \citep{Mccall04}. Taking these considerations into account, we assumed $\Delta E_{B-V}=0.3E_{B-V}$ for all stars. The uncertainties of other colour excesses are slightly larger due to the $\sim10$\% uncertainty in the extinction ratios \citep[see, e.g., ][]{Taylor86}.

\subsubsection{Surface gravity}
    The surface gravity $\log{g}$ (cgs) can be computed from the fundamental relation
\begin{equation}\label{eqn:logg}
\log{g} = \log{g_\sun} + \log{\frac{M}{M_\sun}} + \log{\frac{L_\sun}{L}} + 4\log{\frac{T_\text{eff}}{T_{\text{eff}, \sun}}}
\end{equation}
by deriving $T_\mathrm{eff}$ and luminosity $L$ (from photometry and parallax), and then using theoretical stellar evolutionary tracks to get the mass $M$. Unfortunately, the parallax uncertainties are generally too large to effectively constrain $L$ and $M$ for the stars of this study (see Col.~(5) of Table~\ref{tbl:sample}). Therefore, we used the PARAM tool\footnote{http://stev.oapd.inaf.it/cgi-bin/param\_1.1} \citep{daSilva06} which adopts two Bayesian priors to take into account the fact that the stars are not distributed randomly on the Hertzsprung-Russell diagram: 1)~the masses are distributed according to the lognormal form of \citet{Chabrier01} initial mass function; 2)~the star formation rate has been constant throughout the last 12 Gyr (the maximum age of a star). The theoretical stellar isochrones of \citet{Girardi00} are then used to compute the probability density functions of $M$ and $\log{g}$, from which the mean surface gravity and its uncertainty are computed (see \citet{daSilva06} for a detailed description of the method) and returned to the user. This probabilistic approach allows the derivation of precise surface gravities ($\Delta\log{g}\sim0.1$) even when the parallax errors are large ($\Delta\pi/\pi\sim50$\%). However, we noticed that it breaks down when $\Delta\pi$ approaches and exceeds $\pi$, which is the case for 7 stars of our sample.

\subsubsection{Metallicity}

\begin{table}
\caption{\label{tbl:speclines} Spectral lines of interest.}
\centering
\begin{tabular}{c c c c c}
\hline\hline
$\lambda$, air & Transition & $\chi$ & \multicolumn{2}{c}{$\log{gf}$}\\
(nm) & & (eV) & (Ref.) & (adopted) \\
\hline
\multicolumn{5}{c}{[\ion{S}{i}]} \\
1082.1176 & $^3{\rm P}_2{-}{^1}{\rm D}_2$ & 0.00 & $-8.704$ & $-8.704$ \\ 
\hline
\multicolumn{5}{c}{\ion{Fe}{i}} \\
1081.8277 & $^3{\rm D}_1^\circ{-}{^3}{\rm P}_1$ & 3.96 & $-1.948$ & $-2.212$ \\ 
1083.3964 & $^5{\rm D}_3{-}{^3}{\rm F}_3^\circ$ & 5.59 & $-1.019$ & $-1.240$ \\ 
1084.9462 & $^5{\rm D}_4{-}{^5}{\rm D}_3^\circ$ & 5.54 & $-1.495$ & $-0.650$ \\ 
1086.3519 & $^3{\rm D}_3^\circ{-}{^5}{\rm F}_4$ & 4.73 & $-0.903$ & $-1.000$ \\ 
1088.1759 & $^3{\rm P}_1{-}{^3}{\rm F}_2^\circ$ & 2.85 & $-3.553$ & $-3.383$ \\ 
1088.4267 & $^3{\rm D}_2^\circ{-}{^3}{\rm P}_2$ & 3.93 & $-1.927$ & $-2.075$ \\ 
1089.6300 & $^3{\rm P}_1{-}{^3}{\rm P}_2^\circ$ & 3.07 & $-2.692$ & $-2.822$ \\ 
\hline
\multicolumn{5}{c}{\ion{Cr}{i}} \\
1080.1362 & $^5{\rm D}_1{-}{^5}{\rm D}_2^\circ$ & 3.01 & $-1.715$ & $-1.680$ \\ 
1081.6901 & $^5{\rm D}_2{-}{^5}{\rm D}_2^\circ$ & 3.01 & $-1.957$ & $-1.920$ \\ 
1082.1658 & $^5{\rm D}_3{-}{^5}{\rm D}_2^\circ$ & 3.01 & $-1.678$ & $-1.630$ \\ 
1090.5712 & $^7{\rm P}_2^\circ{-}{^7}{\rm S}_3$ & 3.44 & $-0.647$ & $-0.642$ \\\hline
\end{tabular}
\tablefoot{Wavelengths, excitation energies and oscillator strengths of the lines of interest. The line primarily used for metallicity determination is the \ion{Fe}{i} $\lambda 1081.8$~nm line. The reference is \citet{FF11} for the sulphur line and the Vienna Atomic Line Database \citep[VALD; ][]{VALD} for the others.}
\end{table}

    Since the spectra analysed in this work cover a wavelength range of about 10~nm only, we were forced to derive the metallicity primarily from spectrum synthesis of a single \ion{Fe}{i} line at $\lambda1081.8$~nm. However, for about a half of the stars we were able to use secondary metallicity diagnostics (given in Table~\ref{tbl:speclines}) to verify that the metallicities are not in serious error.

    Synthetic spectra were computed with the Spectroscopy Made Easy \citep[SME;][]{Valenti96} software using the MARCS model atmospheres of \citet{Gustafsson08}. The main assumptions of the MARCS models are spherical symmetry, hydrostatic equilibrium, flux constancy and local thermodynamic equilibrium. Convection is treated using the mixing-length formalism with mixing-length parameter $\alpha=\ell/H_\mathrm{p}=1.5$ \citep[$H_\mathrm{p}$ is the local pressure scale height; ][]{Henyey65}. We used the $\alpha$-enhanced models for which $\text{[}\alpha\text{/Fe]}$ linearly increases from 0 at solar metallicity and above to $+0.4$ at $\mathrm{[Fe/H]}=-1$ and below. The abundances are derived by letting SME do multiple runs, in between which the desired abundances are adjusted, until a satisfactory agreement between synthetic and observed spectra (determined by $\chi^2$ minimization) is achieved. The synthetic spectra are calculated by convolving the theoretical flux spectra with a Gaussian profile to account for macroturbulence and instrumental broadening. The abundances of other $\alpha$-elements in SME were fixed to be consistent with the MARCS models.
            
    We derived astrophysical $\log{gf}$ values for the iron lines by fitting these lines in the solar centre intensity spectrum of \citet{Wallace93} (see Table~\ref{tbl:speclines}). The resulting $\log{gf}$ values were tested by producing a synthetic flux spectrum of Arcturus for which the stellar parameters were taken from \citet{Ramirez11}. The observations \citep{Hinkle95} were well reproduced using the astrophysical $\log{gf}$ values -- for the worst fitting line at $1088.2$~nm the $\log{gf}$ value may be in error by about 0.05~dex, which, when assumed as the uncertainty in the $\log{gf}$ values, would increase the uncertainties in [Fe/H] and [S/Fe] by a further $\sim0.02$ dex. Astrophysical $\log{gf}$ values for four chromium lines in the region were also derived because the [\ion{S}{i}] line is slightly blended by a \ion{Cr}{i} line at $\lambda1081.7$~nm in the coolest/most metal-rich stars of our sample. 
    
    We included the molecular species CN (an up-to-date compilation privately provided by B.~Plez), C$_2$ \citep[][U.~J\o rgensen, private communication]{Querci71} and CH \citep{Jorgensen96} in the computations of synthetic spectra.
        
    In our methodology the metallicity is some function of $T_\mathrm{eff}$ and $\log{g}$, which itself is some complicated function of $T_\mathrm{eff}$. Due to this dependence the uncertainties cannot be added quadratically. But, a conservative estimate of the uncertainties of the metallicities can be obtained by writing the full differential of $\mathrm{[Fe/H]}(T_\mathrm{eff},\log{g})$:
\begin{equation}\label{eqn:feh_spar_err}
\Delta \mathrm{[Fe/H]}_0=\left |\frac{\partial \mathrm{[Fe/H]}}{\partial T_\mathrm{eff}}\Delta T_\mathrm{eff}\right | + \left |\frac{\partial \mathrm{[Fe/H]}}{\partial \log{g}}\Delta \log{g}\right|.
\end{equation}
Here $\Delta T_\mathrm{eff}$ and $\Delta\log{g}$ are given by Eq.~\eqref{eqn:T_meanerr} and PARAM, respectively. An additional uncertainty, mainly due to the uncertainty in continuum placement, is independently introduced by fitting the synthetic profile to the observed profile. This uncertainty was estimated to contribute $\Delta \mathrm{[Fe/H]}_\text{syn}=0.03$ or 0.06~dex depending on the signal-to-noise ratio of the spectrum. The total uncertainty in metallicity was calculated as
\begin{equation}\label{eqn:feh_spar_toterr}
\Delta \mathrm{[Fe/H]}=\sqrt{\left(\Delta \mathrm{[Fe/H]}_0\right)^2+\left(\Delta \mathrm{[Fe/H]}_\text{syn}\right)^2}.
\end{equation}

The two partial derivatives in Eq.\,\eqref{eqn:feh_spar_err} were estimated by deriving the change in metallicity due to changes in temperature (gravity) by $\delta T_\mathrm{eff}=\pm50$\,K and $\delta T_\mathrm{eff}=\pm100$\,K ($\delta \log{g}=\pm0.2$\,dex and $\delta \log{g}=\pm0.4$\,dex). This gives four $\delta \mathrm{[Fe/H]}/\delta T_\mathrm{eff}$ ($\delta \mathrm{[Fe/H]}/\delta \log{g}$) values, one for each 50\,K (0.2\,dex) interval, and the mean of these four values was adopted as the approximation to $\partial \mathrm{[Fe/H]}/\partial T_\mathrm{eff}$ ($\partial \mathrm{[Fe/H]}/\partial \log{g}$) at the given $T_\mathrm{eff}$ ($\log{g}$).

    Since $\mathrm{[Fe/H]}$ has to be known before computing both $T_\mathrm{eff}$ and $\log{g}$, it was first adopted from literature (last column of Table~\ref{tbl:spar_final}; see Sect.~\ref{ch:res}). After the effective temperature and surface gravity were computed, the metallicity of the model atmosphere was adjusted until the best agreement between theoretical and observed spectra was obtained. Since this metallicity was almost always different from the one taken from literature, the stellar parameters were then recalculated, starting with $T_\mathrm{eff}$ and ending with $\mathrm{[Fe/H]}$, until they converged to self-consistent values. One such iteration was typically sufficient to obtain convergence. This procedure lead to an average reduction in the uncertainties of $T_\mathrm{eff}$ and $\log{g}$ by about $10$\%.
    
\subsubsection{Microturbulence}
    There are too few iron lines to constrain the microturbulence from the spectra. However, since the lines of interest are weak, we were not concerned with precise values of $\xi$, and were content with estimating them from the $\xi=\xi(T_\mathrm{eff}, \log{g}, \mathrm{[Fe/H]})$ calibrations privately provided by M.~Bergemann (derived for Gaia-ESO), which are accurate to (i.e., the difference from measured values is) about 0.2--0.3~km\,s$^{-1}$ (standard deviation). A 0.3~km\,s$^{-1}$ error in $\xi$ introduces a negligible error in the sulphur abundance (0.02--0.03~dex) even for stars with the strongest sulphur lines (e.g., HD~65953).

\subsection{Sulphur abundances}\label{ch:meth_sfe}

   The sulphur abundances were derived from the [\ion{S}{i}] line in the same way as the metallicities, once all stellar parameters were determined. The atomic data of the sulphur line were adopted from the compilation of \citet{FF11}.
   
   The uncertainties in $\mathrm{[S/Fe]}$ are
\begin{equation}\label{eqn:sfe_toterr}
\Delta \mathrm{[S/Fe]}=\sqrt{\left(\Delta \mathrm{[S/Fe]}_0\right)^2+\left(\Delta \mathrm{[S/Fe]}_\text{syn}\right)^2},
\end{equation}
where $\mathrm{[S/Fe]}_0$ is the uncertainty due to all stellar parameters (except for $\xi$) and is given by an expression analogous to Eq.~\eqref{eqn:feh_spar_err} with a third term corresponding to $\Delta\mathrm{[Fe/H]}$, and we assume that $\Delta\mathrm{[S/Fe]}_\mathrm{syn}$ is also 0.03 or 0.06~dex (because the iron and sulphur lines are of similar strength) but not necessarily the same as $\Delta\mathrm{[Fe/H]}_\mathrm{syn}$. For most stars $\Delta\mathrm{[S/Fe]}_\mathrm{syn}$ is at least equal to $\Delta\mathrm{[Fe/H]}_\mathrm{syn}$ because generally it is more difficult to place the continuum for the sulphur line due to partial blending with the \ion{Cr}{i} $\lambda 1082$~nm line in the red wing and molecular lines in the blue wing.

\section{Results}\label{ch:res}

\begin{table*}
\caption{\label{tbl:spar_final} Stellar parameters derived in this and previous works.}
\centering
\begin{tabular}{r c c c c c c c c c c c c}
\hline\hline
HD/BD & \multicolumn{9}{c}{This study} & \multicolumn{3}{c}{Literature data} \\
 & $T_\mathrm{eff}$ & $\Delta T_\mathrm{eff}$ & $\log{g}$ & $\Delta\log{g}$ & $\mathrm{[Fe/H]}$ & $\Delta\mathrm{[Fe/H]}$ & $i$ & $\xi$ & [(Fe+Cr)/H] & $T_\mathrm{eff}$ & $\log{g}$ & $\mathrm{[Fe/H]}$ \\
\hline
2796     & $5002$  & $ 30$  & $ 1.73 $ & $0.20$ & $(-2.33)$ & $0.13$ & \ldots & 1.8 & \ldots  & 4932 & 1.40 & $-2.33$   \\
3546     & $4977$  & $ 30$  & $ 2.44 $ & $0.12$ & $ -0.54 $ & $0.04$ & 1, 2 & 1.4 & $-0.64$ & 4923 & 2.48 & $-0.63$  \\
8724     & $4790$  & $ 30$  & $ 1.66 $ & $0.12$ & $ -1.55 $ & $0.07$ & 1  & 1.8 & \ldots  & 4574 & 1.35 & $-1.75$   \\
10380    & $4216$  & $ 30$  & $ 1.48 $ & $0.08$ & $ -0.23 $ & $0.06$ & 1  & 1.4 & $-0.20$ & 4121 & 1.72 & $-0.27$   \\
13979    & $5052$  & $159$  & $(1.50)$ & $0.49$ & $(-2.57)$ & $0.24$ & \ldots & 1.9 & \ldots  & 5043 & 1.50 & $-2.57$   \\
21581    & $5051$  & $ 30$  & $ 2.37 $ & $0.14$ & $ -1.50 $ & $0.04$ & 1, 6, 7 & 1.6 & \ldots  & 4893 & 2.09 & $-1.70$   \\
23798    & $4482$  & $ 33$  & $ 0.71 $ & $0.12$ & $ -2.10 $ & $0.07$ & 1, 5, 6, 7 & 2.1 & \ldots  & 4442 & 0.79 & $-2.12$   \\
26297    & $4484$  & $ 30$  & $ 0.83 $ & $0.10$ & $ -1.72 $ & $0.04$ & 1, 3, 5, 6 & 2.0 & \ldots  & 4432 & 1.13 & $-1.75$   \\
29574    & $4299$  & $ 30$  & $ 0.51 $ & $0.34$ & $ -1.97 $ & $0.06$ & 1, 3, 5, 6 & 2.1 & \ldots  & 4154 & 0.04 & $-1.93$   \\
34334    & $4308$  & $ 30$  & $ 1.87 $ & $0.06$ & $ -0.40 $ & $0.05$ & 1, 3, 4 & 1.4 & $-0.36$ & 4194 & 2.17 & $-0.37$  \\
36702    & $4337$  & $ 30$  & $(0.88)$ & $0.10$ & $ -2.12 $ & $0.07$ & 1, 3 & 2.0 & \ldots  & 4394 & 0.88 & $-2.03$   \\
37160    & $4771$  & $ 30$  & $ 2.57 $ & $0.05$ & $ -0.59 $ & $0.06$ & 1, 3 & 1.3 & $-0.56$ & 4729 & 2.65 & $-0.47$  \\
40460    & $4724$  & $ 45$  & $ 2.42 $ & $0.08$ & $ -0.21 $ & $0.05$ & 1 & 1.2 & $-0.29$ & 4541 & 2.00 & $-0.50$   \\
44007    & $4907$  & $ 30$  & $ 2.36 $ & $0.07$ & $ -1.67 $ & $0.06$ & 1, 3, 6 & 1.6 & \ldots  & 4862 & 2.10 & $-1.64$   \\
65953    & $4038$  & $ 30$  & $ 1.19 $ & $0.07$ & $ -0.30 $ & $0.07$ & 1 & 1.5 & $-0.28$ & 3960 & 1.68 & $-0.36$   \\
81192    & $4779$  & $ 30$  & $ 2.46 $ & $0.04$ & $ -0.71 $ & $0.04$ & 1, 2 & 1.4 & $-0.74$ & 4728 & 2.58 & $-0.64$   \\
83212    & $4616$  & $ 30$  & $ 1.26 $ & $0.10$ & $ -1.44 $ & $0.04$ & 1, 5, 6 & 1.8 & \ldots  & 4518 & 1.38 & $-1.45$   \\
85773    & $4443$  & $ 30$  & $ 1.00 $ & $0.12$ & $(-2.24)$ & $0.24$ & \ldots & 2.0 & \ldots  & 4431 & 0.91 & $-2.24$   \\
103545   & $5085$  & $ 60$  & $(1.70)$ & $0.30$ & $ -1.70 $ & $0.08$ & 1 & 1.8 & \ldots  & 4708 & 1.70 & $-2.14$   \\
110184   & $4374$  & $ 30$  & $(0.65)$ & $0.40$ & $ -2.33 $ & $0.10$ & 1 & 2.1 & \ldots  & 4342 & 0.65 & $-2.33$   \\
111721   & $5078$  & $ 30$  & $ 2.56 $ & $0.17$ & $ -1.30 $ & $0.07$ & 1 & 1.5 & \ldots  & 5038 & 2.70 & $-1.34$   \\
117876   & $4772$  & $ 30$  & $ 2.35 $ & $0.10$ & $ -0.44 $ & $0.04$ & 1, 2 & 1.3 & $-0.41$ & 4722 & 2.27 & $-0.49$   \\
122563   & $4719$  & $ 30$  & $ 1.32 $ & $0.04$ & $(-2.65)$ & $0.16$ & \ldots & 1.9 & \ldots  & 4594 & 1.25 & $-2.65$   \\
122956   & $4698$  & $ 37$  & $ 1.44 $ & $0.11$ & $ -1.74 $ & $0.07$ & 1 & 1.8 & \ldots  & 4618 & 1.50 & $-1.74$   \\
139195   & $4994$  & $ 66$  & $ 2.85 $ & $0.05$ & $  0.00 $ & $0.05$ & 1, 2 & 1.1 & $-0.04$ & 4991 & 3.03 & $-0.11$  \\
161074   & $3949$  & $ 30$  & $ 1.45 $ & $0.07$ & $ -0.03 $ & $0.09$ & 1 & 1.3 & $-0.05$ & 3997 & 1.83 & $-0.27$   \\
166161   & $5195$  & $ 38$  & $ 2.52 $ & $0.13$ & $ -1.18 $ & $0.07$ & 1 & 1.5 & \ldots  & 5177 & 2.23 & $-1.21$   \\
168723   & $4936$  & $ 30$  & $ 3.04 $ & $0.03$ & $ -0.22 $ & $0.04$ & 1 & 1.1 & $-0.21$ & 4902 & 3.09 & $-0.14$  \\
184406   & $4530$  & $ 30$  & $ 2.70 $ & $0.06$ & $  0.20 $ & $0.04$ & 1 & 1.0 & $ 0.18$ & 4486 & 2.47 & $-0.04$  \\
187111   & $4418$  & $ 30$  & $ 0.70 $ & $0.11$ & $ -1.78 $ & $0.04$ & 1, 3 & 2.1 & \ldots  & 4280 & 0.76 & $-1.75$   \\
188512   & $5111$  & $ 79$  & $ 3.52 $ & $0.06$ & $ -0.23 $ & $0.06$ & 1, 2 & 1.2 & $-0.21$ & 5107 & 3.56 & $-0.16$  \\
204543   & $4490$  & $ 30$  & $(1.25)$ & $0.30$ & $ -1.97 $ & $0.07$ & 1 & 1.9 & \ldots  & 4682 & 1.25 & $-1.82$  \\
212943   & $4694$  & $ 30$  & $ 2.61 $ & $0.06$ & $ -0.26 $ & $0.04$ & 1, 2 & 1.2 & $-0.25$ & 4611 & 2.79 & $-0.29$  \\
214567   & $4907$  & $103$  & $ 2.57 $ & $0.14$ & $ -0.19 $ & $0.08$ & 1, 2 & 1.2 & $-0.20$ & 5003 & 2.50 & $-0.03$   \\
216143   & $4362$  & $113$  & $(1.05)$ & $0.49$ & $ -2.22 $ & $0.14$ & 1 & 2.0 & \ldots  & 4511 & 1.05 & $-2.22$   \\
219615   & $4843$  & $ 30$  & $ 2.39 $ & $0.08$ & $ -0.56 $ & $0.04$ & 1 & 1.4 & $-0.53$ & 4848 & 2.52 & $-0.49$  \\
220954   & $4766$  & $ 51$  & $ 2.70 $ & $0.10$ & $  0.25 $ & $0.05$ & 1 & 1.0 & $ 0.20$ & 4708 & 2.74 & $-0.03$  \\
221170   & $4591$  & $ 32$  & $ 1.05 $ & $0.10$ & $ -2.00 $ & $0.07$ & 1 & 2.0 & \ldots  & 4480 & 0.98 & $-2.11$   \\
+30~2611 & $4322$  & $ 30$  & $(0.90)$ & $0.33$ & $ -1.43 $ & $0.07$ & 1 & 1.9 & $-1.41$ & 4291 & 0.90 & $-1.40$   \\
\hline
\end{tabular}
\tablefoot{The quantities in parentheses denote values that were adopted from literature (i.e., the last two columns). $i$ denotes the \ion{Fe}{i} lines used in metallicity determination (ordered according to wavelength as in Table~\ref{tbl:speclines}). $\Delta\xi$ is about 0.2--0.3~km\,s$^{-1}$. Literature data are the interquartile mean values from \citet{Soubiran10}.}
\end{table*}

   The stellar parameters derived in this work are given in Table~\ref{tbl:spar_final}. For comparison, we also list the literature values, which are the interquartile mean (IQM) values of the entries in the PASTEL catalogue of stellar parameters \citep{Soubiran10}.\footnote{Whenever a single value of one of the stellar parameters was referenced multiple times, all but one of duplicate values from different studies by the same first author were discarded to reduce the influence of a single study.}

   \begin{figure}
   \centering
   \includegraphics[width=\hsize]{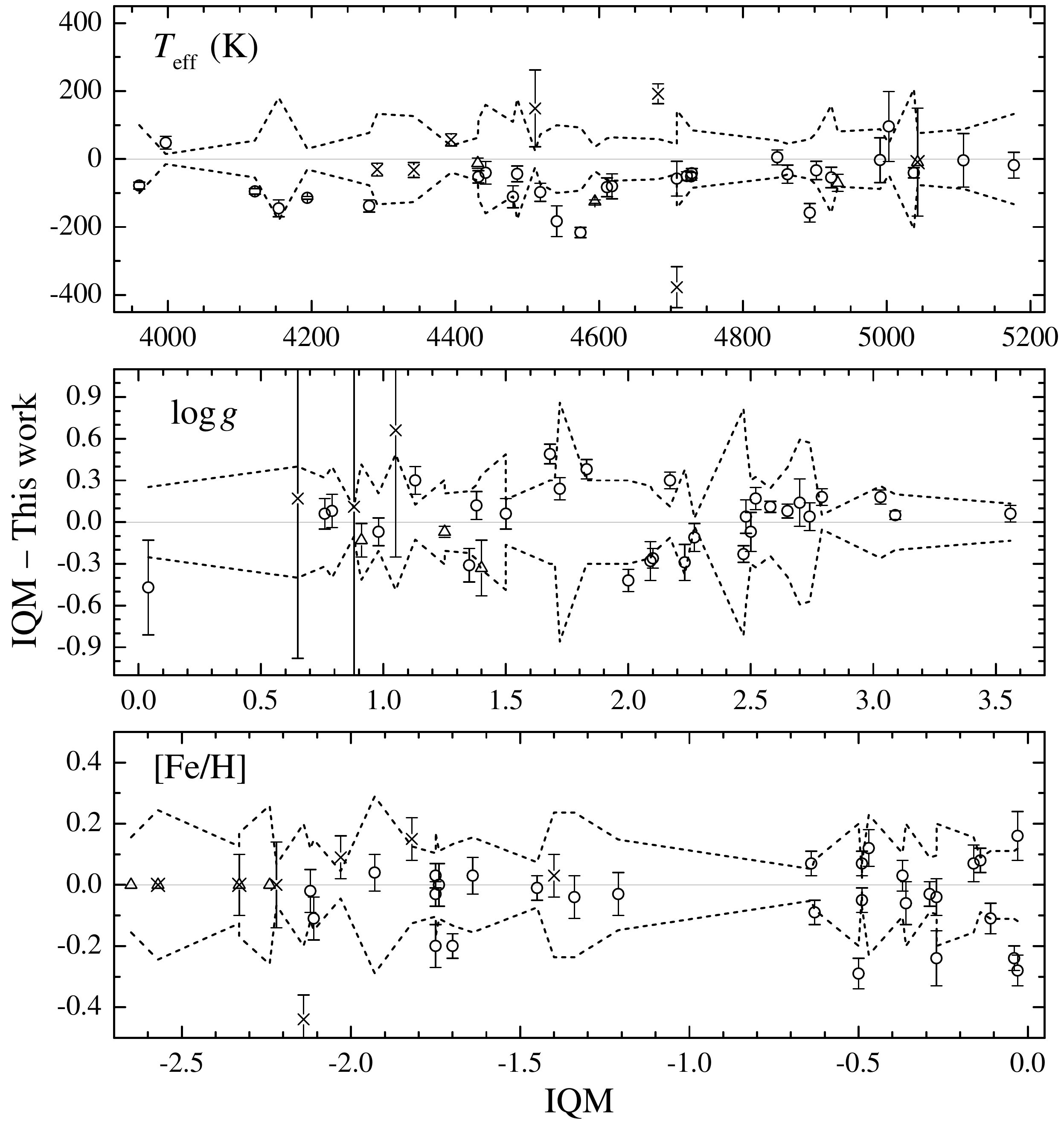}
      \caption{Differences between stellar parameters from literature \citep[interquartile mean (IQM) values from][]{Soubiran10} and the stellar parameters derived in this study versus the literature parameters -- top: effective temperatures; middle: surface gravities; bottom: metallicities. The circles correspond to stars with homogeneous stellar parameters, crosses (triangles) denote stars, whose gravities (metallicities) were adopted from literature. The error bars are given by Eq.~\eqref{eqn:T_meanerr}, PARAM (see text) and Eq.~\eqref{eqn:feh_spar_toterr} for the three respective stellar parameters. The jagged line indicates the uncertainty of literature values (IQR/1.35).}
         \label{fig:spar}
   \end{figure}

   All four stellar parameters have been derived for 29 out of the 39 stars. For the remaining ten stars, all of which have $\mathrm{[Fe/H]}<-1$, the IQM values of either the surface gravity or the metallicity (or both) were used. For 7 of the 10 stars the parallax errors exceed the measured parallax, and so the surface gravity could not be derived. The metallicity could not be derived for 4 of the 10 stars, because none of the iron lines were detected in their spectra. 
      
   Our sample covers effective temperatures from 3900 to 5200~K, surface gravities mainly from 0.5 to 3.0, and metallicities mainly from $-2.4$ to $+0.2$. In general the formally propagated temperature uncertainties are very small (between 30 and 50~K), which indicates that the individual colours give consistent estimates of $T_\mathrm{eff}$. Larger uncertainties are associated with fewer colours used or could indicate problems with reddening corrections. The lower limit of 30~K is artificial -- for most stars Eq.~\eqref{eqn:T_meanerr} gives a smaller uncertainty of the effective temperature, which can make the derived surface gravities very sensitive to the exact value of this uncertainty. Therefore, for every star $\Delta T_\mathrm{eff}$ was adjusted to equal the result of Eq.~\eqref{eqn:T_meanerr} or 30~K, whichever is larger, before calculating its surface gravity, metallicity and sulphur abundance. $\Delta\log{g}$ are typically 0.1--0.2~dex. Larger values ($\sim0.4$~dex) are characteristic for the 7 stars with IQM gravities (their uncertainties are calculated as the interquartile range (IQR) divided by 1.35, which would equal 1-$\sigma$ uncertainties in case of normally distributed values). Uncertainties in $\mathrm{[Fe/H]}$ are normally below 0.1~dex, but again larger values are typical for the 4 stars with IQM metallicities.

    In abundance studies of low-mass main-sequence stars a very tight correlation between the chromium and iron abundances is observed \citep{Bensby05}, with $\mathrm{[Cr/H]}\simeq\mathrm{[Fe/H]}$ in stars with $\mathrm{[Fe/H]}>-1$. Under the reasonable assumption that this correlation extends to giants, we have calculated ``updated'' metallicities by treating the chromium abundance as another measure of the metallicity.\footnote{Logarithm of the number of chromium atoms per $10^{12}$ hydrogen atoms in the sun: $\log\epsilon(\mathrm{Cr})_\sun=5.64$ \citep{Grevesse07}.} The updated values are given in Col.~(9) of Table~\ref{tbl:spar_final}. As evidenced by the very small changes in the metallicity, the Cr and Fe abundances are indeed nearly equal for the most part. However, to remain consistent throughout the metallicity range, the updated metallicities are not considered further.
    
   The stellar parameters derived in this study are plotted against their respective IQM values in Fig.~\ref{fig:spar}. Open circles are used for stars with homogeneous stellar parameters, crosses for stars with IQM surface gravities and triangles for stars with IQM metallicities. The jagged dashed line denotes the uncertainty in the literature values, which is equal to the IQR value divided by 1.35. The error bars are calculated from Eq.~\eqref{eqn:T_meanerr} in the case of $T_\mathrm{eff}$ (top panel; notice that some are clearly smaller than the minimum adopted $\Delta T_\mathrm{eff}$ of 30~K), returned by PARAM in case of $\log{g}$ (middle panel), and calculated from Eq.~\eqref{eqn:feh_spar_toterr} in case of $\mathrm{[Fe/H]}$ (bottom panel).

   \begin{figure*}
   \sidecaption
   \includegraphics[width=12cm]{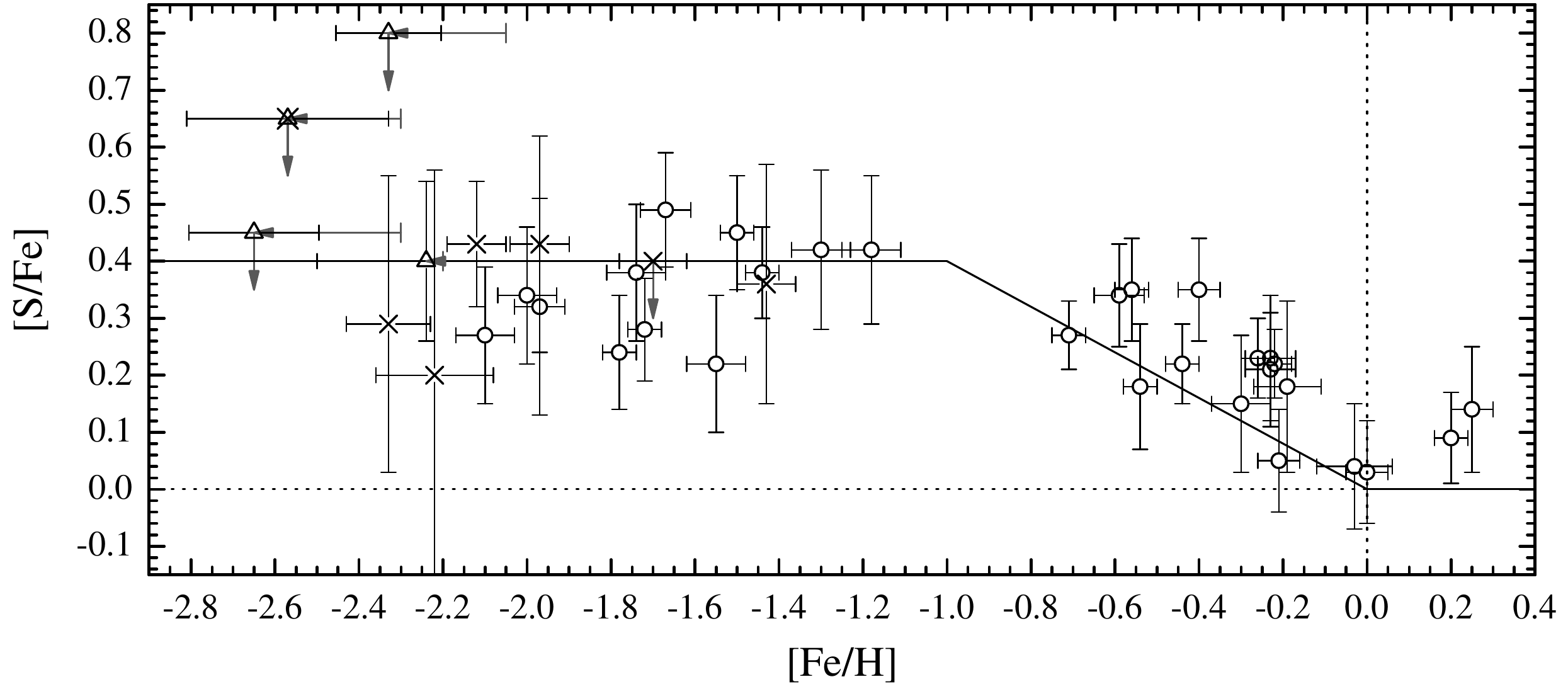}
      \caption{Sulphur abundances derived in this work. The symbols are the same as in Fig.~\ref{fig:spar}. The starting points of the horizontal (vertical) arrows indicate the upper limit of the metallicity (sulphur abundance). The (upper limit of) sulphur abundance for these stars is derived with respect to the literature (IQM) metallicity. The solid line shows the enrichment assumed for the other $\alpha$-elements. The dotted lines cross at the solar values, which correspond to $\log\epsilon(\mathrm{Fe})_\sun=7.45$ and $\log\epsilon(\mathrm{S})_\sun=7.14$ \citep{Grevesse07}.}
         \label{fig:sferes}
   \end{figure*}

\begin{figure*}
   \centering
   \includegraphics[width=\hsize]{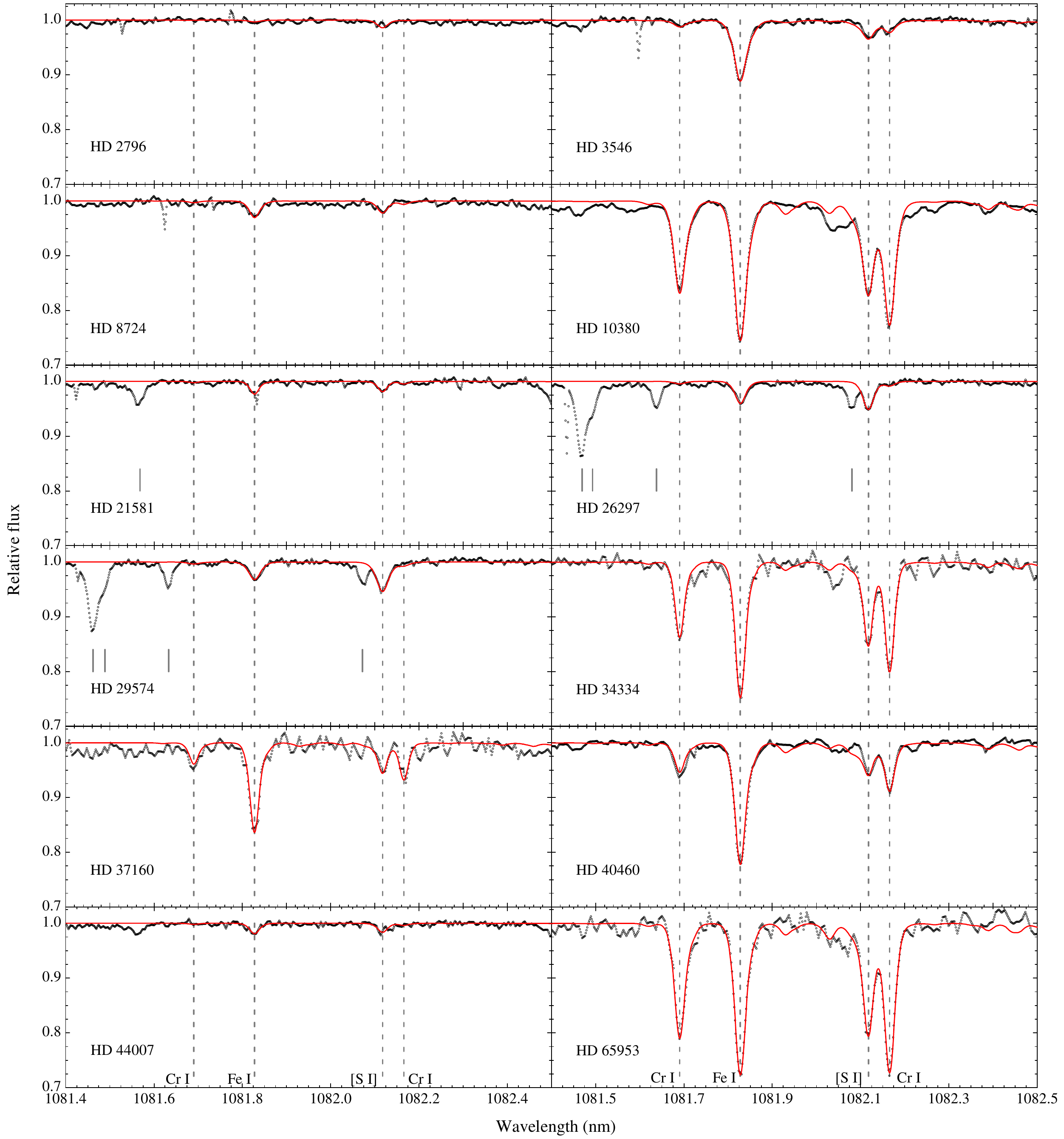}
      \caption{Synthetic spectra (red line) plotted over the observations (black circles) for a sample of the stars. The positions of important lines are marked by vertical dashed lines. The vertical ticks mark the positions of telluric lines.}
         \label{fig:synsfe1}
   \end{figure*}

\begin{figure*}
   \centering
   \includegraphics[width=\hsize]{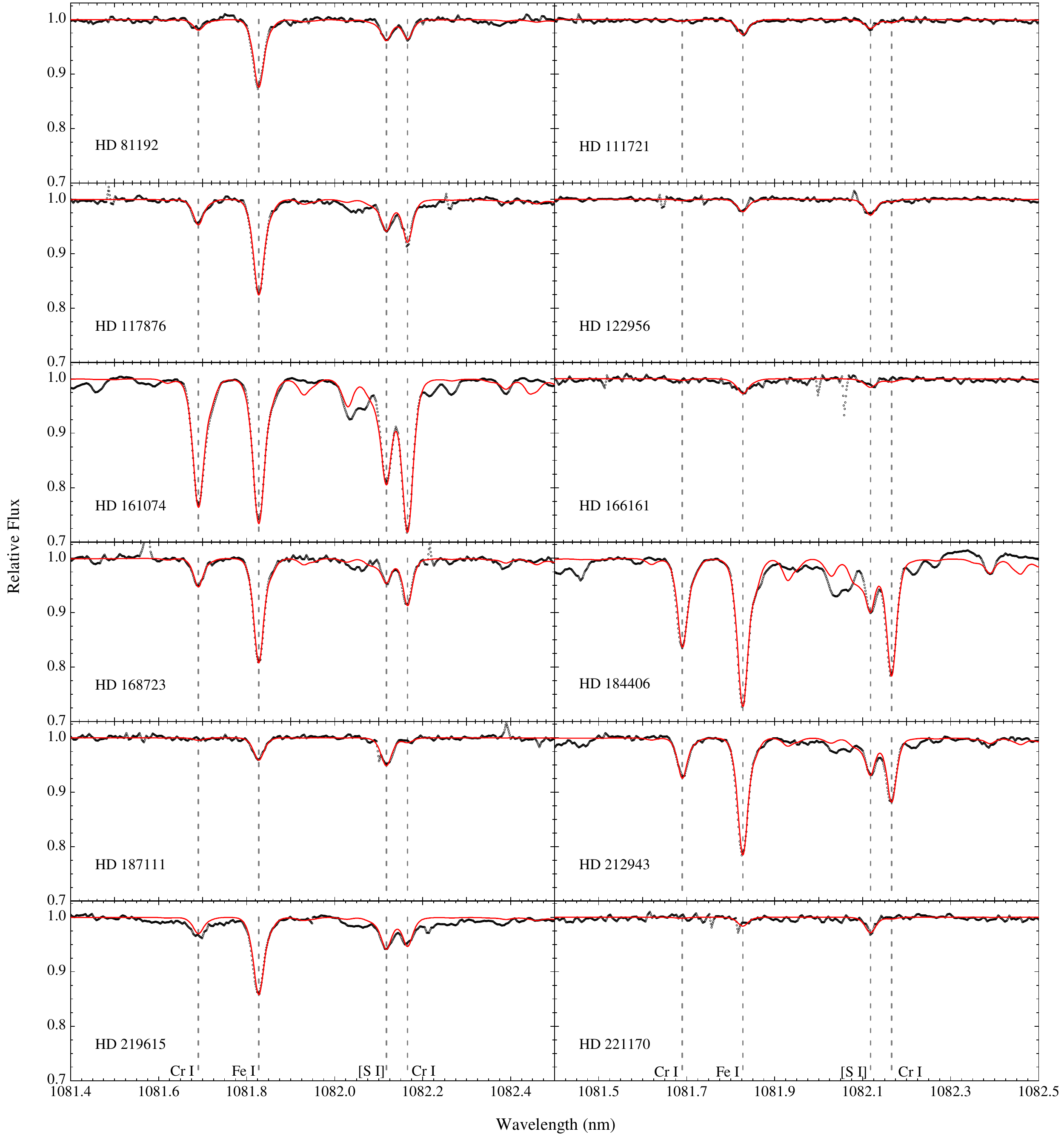}
      \caption{Synthetic spectra (red line) plotted over the observations (black circles) for a sample of the stars. The positions of important lines are marked by vertical dashed lines.}
         \label{fig:synsfe2}
   \end{figure*}

   The sulphur abundances derived in this work are listed in the third column of Table~\ref{tbl:prevsfe} and plotted versus metallicities in Fig.~\ref{fig:sferes}. The symbols are the same as in Fig.~\ref{fig:spar}. The starting points of the horizontal (vertical) arrows indicate the upper limits of the metallicities (sulphur abundances). For reference, the abundance of the other $\alpha$-elements adopted for spectrum synthesis is denoted by the solid line. The lack of stars with $-1.2<\mathrm{[Fe/H]}<-0.8$ is a chance outcome. While typical uncertainties of $\mathrm{[S/Fe]}$ are 0.1--0.15~dex, $\Delta\mathrm{[S/Fe]}$ can easily reach $\sim0.4$~dex when the stellar parameters are poorly constrained. Some examples of the theoretical spectra (red lines) fitted to observations (black circles) are presented in Figs.~\ref{fig:synsfe1} and \ref{fig:synsfe2}, in which a 1~nm wide spectral region containing the forbidden sulphur line is shown. This region also contains the $\lambda1081.8$~nm iron line used to derive metallicities and two of the four chromium lines listed in Table~\ref{tbl:speclines}. The positions of these spectral lines are indicated by vertical dashed lines.

\begin{table}
\caption{\label{tbl:prevsfe} The sulphur abundances derived in this and previous works.}
\centering
\begin{tabular}{r c c c c c}
\hline\hline
HD/BD & \multicolumn{3}{c}{This study} & \multicolumn{2}{c}{Previous studies} \\
 & $\mathrm{[Fe/H]}$ & $\mathrm{[S/Fe]}$ & $\Delta\mathrm{[S/Fe]}$ & $\mathrm{[Fe/H]}$ & $\mathrm{[S/Fe]}$ \\
\hline
8724   & $-1.55$ & 0.22 & 0.12 & \ldots & \ldots \\ 
34334  & $-0.40$ & 0.35 & 0.09 & \ldots & \ldots \\ 
37160  & $-0.59$ & 0.34 & 0.09 & \ldots & \ldots \\ 
65953  & $-0.30$ & 0.15 & 0.12 & \ldots & \ldots \\ 
110184 & $-2.33$ & 0.29 & 0.26 & \ldots & \ldots \\ 
122956 & $-1.74$ & 0.38 & 0.12 & \ldots & \ldots \\ 
166161 & $-1.18$ & 0.42 & 0.13 & \ldots & \ldots \\ 
204543 & $-1.97$ & 0.43 & 0.19 & \ldots & \ldots \\ 
+302611 & $-1.43$ & 0.36 & 0.21 & \ldots & \ldots \\ 
\hline
\multicolumn{6}{c}{\citet{Ryde04}}\\
111721 &           $-1.30$ & $0.42$ & 0.14      &            $-1.21$ & $\phantom{-}0.32$  \\
\hline
\multicolumn{6}{c}{\citet{TakH05}}\\
187111 &            $-1.78$ & $0.24$ & 0.10  &           $-1.85$ & $\phantom{-}0.66$  \\
216143 &            $-2.22$ & $0.20$ & 0.36  &           $-2.15$ & $\phantom{-}0.37$  \\
221170 &            $-2.00$ & $0.34$ & 0.12  &           $-2.10$ & $\phantom{-}0.47$  \\
\hline
\multicolumn{6}{c}{\citet{Ryde06}}\\
3546   &            $-0.54$ & $0.18$ & 0.11  &           $-0.60$ & $\phantom{-}0.25$  \\
10380  &            $-0.23$ & $0.21$ & 0.10  &           $-0.19$ & $\phantom{-}0.03$  \\
40460  &            $-0.21$ & $0.05$ & 0.09  &           $-0.44$ & $\phantom{-}0.00$  \\
81192  &            $-0.71$ & $0.27$ & 0.06  &           $-0.56$ & $\phantom{-}0.15$  \\
117876 &            $-0.44$ & $0.22$ & 0.07  &           $-0.44$ & $\phantom{-}0.10$  \\
139195 &  $\phantom{-}0.00$ & $0.03$ & 0.09  & $\phantom{-}0.02$ &           $-0.14$  \\
161074 &            $-0.03$ & $0.04$ & 0.11  &           $-0.21$ & $\phantom{-}0.20$  \\
168723 &            $-0.22$ & $0.22$ & 0.06  &           $-0.13$ & $\phantom{-}0.14$  \\
184406 &  $\phantom{-}0.20$ & $0.09$ & 0.08  & $\phantom{-}0.07$ &           $-0.06$  \\
188512 &            $-0.23$ & $0.23$ & 0.11  &           $-0.11$ & $\phantom{-}0.20$  \\
212943 &            $-0.26$ & $0.23$ & 0.07  &           $-0.28$ & $\phantom{-}0.21$  \\
214567 &            $-0.19$ & $0.18$ & 0.15  & $\phantom{-}0.09$ &           $-0.08$  \\
219615 &            $-0.56$ & $0.35$ & 0.09  &           $-0.36$ & $\phantom{-}0.24$  \\
220954 &  $\phantom{-}0.25$ & $0.14$ & 0.11  &           $-0.04$ & $\phantom{-}0.09$  \\
\hline
\multicolumn{6}{c}{\citet{Spite11}}\\
2796   &            $(-2.33)$ & $\lesssim0.80\phantom{\lesssim}$ & -- &           $-2.41$ & $\phantom{-}0.31$  \\
122563 &            $(-2.65)$ & $\lesssim0.45\phantom{\lesssim}$ & -- &           $-2.76$ & $\phantom{-}0.38$  \\
\hline
\multicolumn{6}{c}{\citet{Jonsson11}} \\
13979  & $(-2.57)$ & $\lesssim0.65\phantom{\lesssim}$ & --  &  $-2.26$ & $\lesssim0.21$     \\
21581  & $-1.50$ & $0.45$ & 0.10                            &  $-1.64$ & $\phantom{-}0.52$    \\
23798  & $-2.10$ & $0.27$ & 0.12                            &  $-2.03$ & $\phantom{-}0.37$    \\
26297  & $-1.72$ & $0.28$ & 0.09                            &  $-1.51$ & $\phantom{-}0.41$    \\
29574  & $-1.97$ & $0.32$ & 0.19                            &  $-1.70$ & $\phantom{-}0.27$    \\
36702  & $-2.12$ & $0.43$ & 0.11                            &  $-2.06$ & $\phantom{-}0.45$    \\
44007  & $-1.67$ & $0.49$ & 0.10                            &  $-1.65$ & $\phantom{-}0.37$    \\
83212  & $-1.44$ & $0.38$ & 0.08                            &  $-1.40$ & $\phantom{-}0.41$    \\
85773  & $(-2.24)$ & $0.40$ & 0.14                          &  $-2.36$ & $\phantom{-}0.53$    \\
103545 & $-1.70$ & $\lesssim0.40\phantom{\lesssim}$ & --    &  $-2.14$ & $\lesssim0.52$     \\
\hline
\end{tabular}
\tablefoot{The quantities in parentheses denote values that were adopted from literature (last column of Table~\ref{tbl:spar_final}). The metallicities and abundances of previous studies have been rescaled to correspond to $\log\epsilon(\mathrm{Fe})_\sun=7.45$ and $\log\epsilon(\mathrm{S})_\sun=7.14$ \citep{Grevesse07}.}
\end{table}


\section{Discussion}

In this section we discuss some points concerning our stellar parameters (Sect.~\ref{ch:disc_spar}) and sulphur abundances (Sect.~\ref{ch:disc_sfe}), and then compare our results to previous studies on the Galactic chemical evolution of sulphur (Sect.~\ref{ch:disc_sfe_prev}), and end with comparing our results with theoretical expectations (Sect.~\ref{ch:disc_sfe_gce}).

\subsection{Stellar parameters}\label{ch:disc_spar}

    We find no $T_\mathrm{eff}$-dependent differences between the effective temperatures derived from the \citet{GHB09} calibrations and the interquartile mean effective temperatures. However, the former are systematically hotter by $65\pm66~\mathrm{K}$ (standard deviation). We also derived the effective temperatures from the calibrations of \citet{RM05b} and found no statistically significant differences from the IQM temperatures on average. We opted to use the \citet{GHB09} calibrations for subsequent work anyway because the metallicity binning for each colour used by \citet{RM05b} results in the calibrated temperatures being discontinuous across the edges of the bins, with a change as small as 0.01~dex in $\mathrm{[Fe/H]}$ potentially introducing a change of $\sim30$~K in $T_\mathrm{eff}$ for some colours.

    The surface gravities and metallicities are found to very closely (to within 0.05~dex) agree on average with the IQM gravities and metallicities.

   No non-LTE corrections have been applied to the metallicities derived in this work because we are not aware of any non-LTE calculations of \ion{Fe}{i} lines with wavelengths $\lambda>1000~\mathrm{nm}$.

\subsection{Sulphur abundances}\label{ch:disc_sfe}

    We plot the sulphur abundances versus metallicities of the whole sample in Fig.\,\ref{fig:sferes}. Unlike in previous studies by \citet{Israelian01}, \citet{TakH02}, \citet{Caffau05} and \citet{Caffau10}, no stars with extremely high sulphur abundances are observed, nor are the sulphur abundances found to increase towards lower metallicities. In fact, the opposite is observed -- a linear fit for the metal-poor ($\mathrm{[Fe/H]}<-1$) stars with homogeneous stellar parameters gives $\mathrm{[S/Fe]}=0.17\mathrm{[Fe/H]}+0.63$ -- although the significance is low (the standard error of the slope is 0.15) and, when the stars with IQM gravities are included, the trend becomes shallower and insignificant (the slope becomes $0.07\pm0.12$). If the slope is fixed at 0, we get an average abundance of $\mathrm{[S/Fe]}=0.35\pm0.09$ (standard deviation) for $-2.4<\mathrm{[Fe/H]}<-1.1$, which is slightly below the $\mathrm{[}\alpha\mathrm{/Fe]}=0.4$ of the MARCS models. For $-0.8<\mathrm{[Fe/H]}<0.0$ $\mathrm{[S/Fe]}$ is found to decrease as the metallicity increases, which is expected of $\alpha$-elements (solid line). But note that in reality more complicated patterns are observed than assumed in the $\alpha$-enhanced MARCS models -- above $\mathrm{[Fe/H]}\simeq-0.8$ two distinct trends, corresponding to the Thin and Thick disks, are observed \citep{Bensby05,Fuhrmann08}. According to the models of \citet{Kobayashi11} the differences for the sulphur trends are, however, less than 0.05 dex.
    
   The uncertainty introduced in the final results ($\mathrm{[S/Fe]}$) is likely to be large when IQM values of stellar parameters are used. However, the use of IQM values might still be preferred to adopting the values from individual studies unless the missing parameters can be adopted from a single study.

   Our formally propagated uncertainties of $\mathrm{[S/Fe]}$ are comparable to what is usually simply stated in the literature: $\Delta\mathrm{[S/Fe]}\sim0.10$--0.15\,dex. Nevertheless, a detailed breakdown of individual contributions has allowed us to establish that $\log{g}$ and $\mathrm{[Fe/H]}$ are the largest contributors to the uncertainty of $\mathrm{[S/Fe]}$, each introducing a partial error of about 0.05--0.10~dex for most stars. While the forbidden line exhibits low temperature sensitivity (the partial error due to $\Delta T_\mathrm{eff}$ is only about 0.02~dex), our methodology does not take full advantage of this fact, because  we use the temperature-sensitive \ion{Fe}{i} lines to derive the metallicity. Using lines of singly-ionized iron instead would address this flaw and also reduce the possibility of deviations from LTE affecting the results \citep{Mashonkina11,Lind12}. Unfortunately, no suitable \ion{Fe}{ii} lines are located in the spectral region around the sulphur line. Overall, we find that the uncertainties in stellar parameters usually dominate the total uncertainty of $\mathrm{[S/Fe]}$.

   Obviously, deriving abundances from a single sulphur line makes each individual of our results more vulnerable to errors, especially at the lowest ($\mathrm{[Fe/H]}\lesssim-2$) metallicities where the risk of mistaking noise for the iron or sulphur line is increased, and at the highest metallicities because of the possibility of unaccounted weak blends. As stated before, to avoid severe errors in our metallicity estimates, whenever possible we checked whether the metallicities derived from the \ion{Fe}{i} $\lambda1081.8$\,nm line are consistent with other, less prominent iron lines in the region.
   
\subsubsection{Comparison with previous studies}\label{ch:disc_sfe_prev}

   \begin{figure*}
   \sidecaption
   \includegraphics[width=12cm]{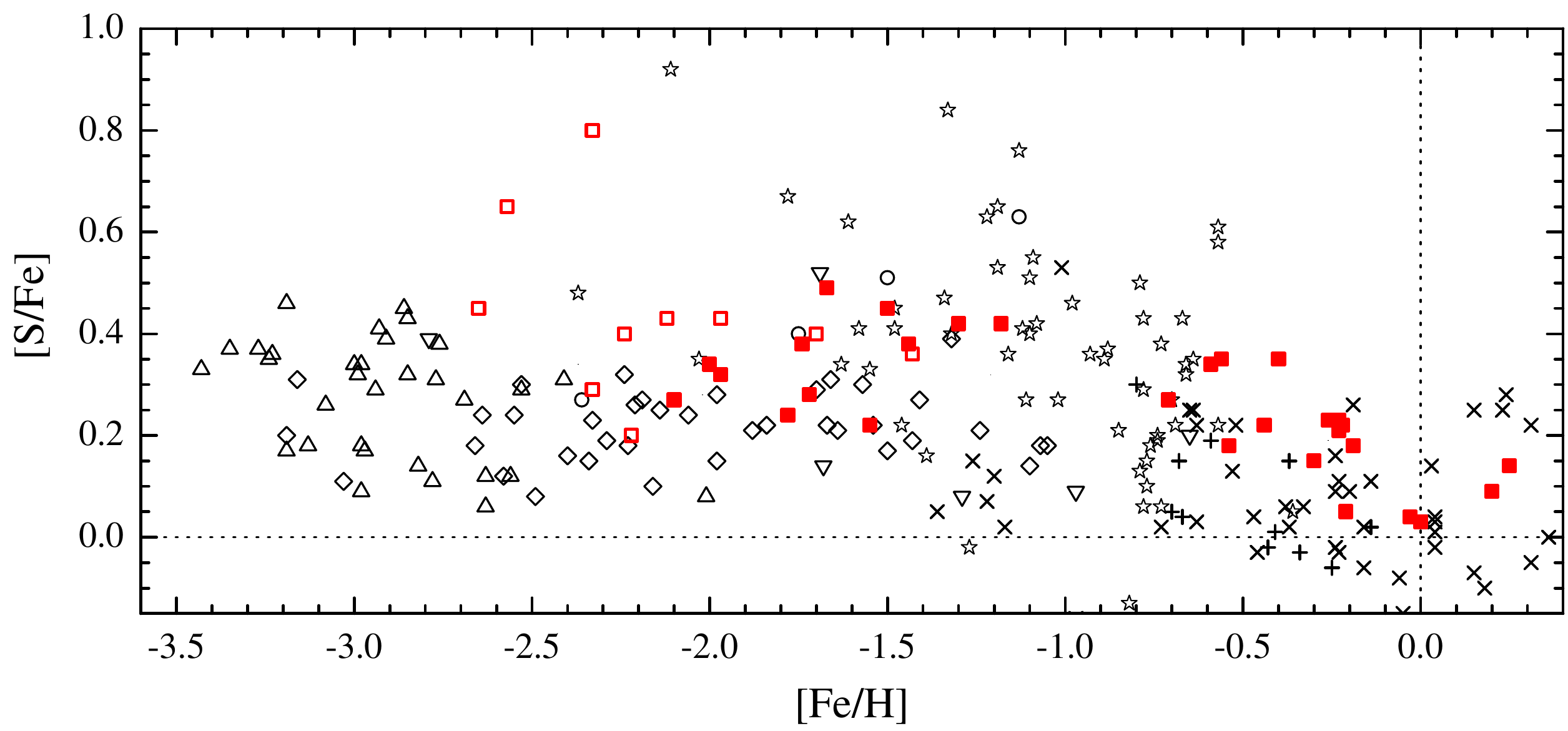}
      \caption{Sulphur abundances derived in this and previous works. Our results are plotted as red squares (solid for stars with homogeneously determined stellar parameters). The plus symbols are measurements from \citet{Chen02}, crosses are from \citet{TakH02}, downwards pointing triangles from \citet{Ryde04}, stars from \citet{Caffau05}, rhombi from \citet{Nissen07}, circles from \citet{Caffau10} and upwards pointing triangles from \citet{Spite11}.}
         \label{fig:sfeprev}
   \end{figure*}

In addition to the \citet{Ryde06} and \citet{Jonsson11} samples, we have some stars in common with \citet{Ryde04}, \citet{TakH05} and \citet{Spite11}, all of who derived the sulphur abundances from the \ion{S}{i} $\lambda923$\,nm lines -- \citet{TakH05} used equivalent widths, and \citet{Ryde04} and \citet{Spite11} performed spectrum synthesis. We list our results along previous results in Table~\ref{tbl:prevsfe}. The metallicities and sulphur abundances of previous studies have been translated to the solar chemical composition scale of \citet{Grevesse07}: $\log\epsilon(\mathrm{Fe})_\sun=7.45$ and $\log\epsilon(\mathrm{S})_\sun=7.14$. This scale was adopted to be consistent with the MARCS models. Adoption of the solar abundances from the more recent study of \citet{Asplund09} would shift all metallicities and sulphur abundances by $-0.05$~dex and $+0.07$~dex, respectively.

    Clearly, there can be major differences in the metallicities and sulphur abundances for individual stars analysed in separate studies. We tested to what extent these differences can be explained by the differences in the stellar parameters. This is particularly important in cases of \citet{Ryde06} and \citet{Jonsson11}, because these studies analysed the same observational data and used the same spectral diagnostics as this study. And indeed, we find that the differences in $\mathrm{[S/Fe]}$ can be explained to within $\sim0.05$\,dex once the differences in stellar parameters are accounted for. In other words, we can reproduce their results if we adopt the same stellar parameters. The answer to the question of which sulphur abundances are the ``correct ones'' therefore depends on which stellar parameters one trusts more. Given the significant differences from study to study, the necessity for homogeneous stellar parameters is apparent. Nevertheless, note that systematic differences between stellar parameters from multiple literature sources may not have a noticeable effect on the main conclusions. For example, \citet{Jonsson11} find a plateau in [S/Fe] with a scatter (standard deviation) of 0.085 dex. We get virtually the same scatter (0.080 dex) for their stars.
    
    Accounting for the differences in stellar parameters is not always enough when comparing the results to the other studies. For example, consider HD~187111, for which we derive a lower $\mathrm{[S/Fe]}$ (by 0.42~dex) than \citet{TakH05} when adopting a higher $T_\mathrm{eff}$ (by 100~K), lower gravity (by 0.4~dex) and slightly higher metallicity (by 0.07~dex). Differences in stellar parameters account for about a half of the discrepancy in $\mathrm{[S/Fe]}$, leaving a difference of about $0.2~\mathrm{dex}$. This remaining difference can be well explained by the strong non-LTE effects of the triplet -- the calculations by \citet{Takeda05} indicate that the result of \citet{TakH05} should be lowered by 0.15~dex. Thus, we argue that HD~187111 has a much lower sulphur abundance than reported in \citet{TakH05}.
    
    In some other cases the differences cannot be explained so easily, and remaining differences, which can be around 0.2~dex in $\mathrm{[S/Fe]}$, must be due to a combination of a number of factors, for example: 1)~quality of observations; 2)~accuracy of continuum normalization; 3)~completeness and accuracy of atomic and molecular data; 4)~uncertainty in the non-LTE corrections; 5)~differences in 3D effects; 6)~different software used for computing synthetic spectra; 7)~different versions of the model atmospheres.
    
   Our mean sulphur abundance below $\mathrm{[Fe/H]}=-1$ ($\mathrm{[S/Fe]}\simeq0.35$) is significantly ($\sim2-3\sigma$) higher than found in previous studies by \citet{Nissen07} and \citet{Spite11}, who find mean ${\mathrm{[S/Fe]}}\simeq0.22$ and $0.27$\,dex, respectively (see Fig.~\ref{fig:sfeprev}). There are three key differences between the studies: 1)~the methods used to obtain the stellar parameters in the three studies are different. \citet{Nissen07} derived $T_\mathrm{eff}$ from the H$\beta$ line, $\log{g}$ from Eq.~\eqref{eqn:logg} and $\mathrm{[Fe/H]}$ from equivalent widths of \ion{Fe}{ii} lines. \citet{Spite11} adopted stellar parameters primarily from three different studies, in which $T_\mathrm{eff}$ was derived from photometric colours and $\log{g}$ from ionization equilibrium; 2)~we use the forbidden line in our analysis, while \citet{Nissen07} and \citet{Spite11} use the \ion{S}{i} triplet at $923$\,nm; 3)~this sample and that of \citet{Spite11} consists mainly of giants and subgiants, while the sample of \citet{Nissen07} mainly comprises dwarf stars. A systematic difference between giants and dwarfs would be a puzzle to solve for model atmospheres. While such a difference may exist, the sulphur abundances derived by \citet{Spite11} for the few dwarf stars in their sample show good agreement with their results for giants, which argues against this possibility. Systematic differences between the stellar parameter scales and sulphur diagnostics, likely in the form of non-LTE and 3D corrections, are more probable. Calculations by \citet{Takeda05} and \citet{Korotin09} have shown that the non-LTE corrections are extremely dependent on temperature and metallicity for metal-poor stars (e.g., the non-LTE corrections for stars with $T_\mathrm{eff}=5000$ and $5500$\,K at $\mathrm{[Fe/H]}=-2$ are $\Delta\mathrm{[S/Fe]}\simeq-0.4$ and $\Delta\mathrm{[S/Fe]}\simeq-0.6$\,dex, respectively, and at $\mathrm{[Fe/H]}=-3$ they are $\Delta\mathrm{[S/Fe]}\simeq-0.4$ and $\Delta\mathrm{[S/Fe]}\simeq-0.8$\,dex, respectively). It is possible that the non-LTE corrections for the triplet are systematically too large by about 0.1\,dex for these types of stars. On a suggestive side note, \citet{Jonsson11} found their non-LTE-corrected sulphur abundances from the triplet at $\lambda1045$\,nm to be lower by an average of $0.06$\,dex (compared to the abundances derived from the forbidden line) in stars of subsample 3. On the other hand, while we emphasize again that the [\ion{S}{i}] line forms in LTE, deviations from LTE are still likely important for our sulphur abundances because of their effects on the \ion{Fe}{i} lines that were used to derive the metallicities. As stated before, the non-LTE effects of the iron lines with $\lambda>1000$~nm have not been studied. 

Also remember that no attempts have been made to account for 3D effects on the results of this study. \citet{Jonsson11} used the 3D hydrodynamic model atmospheres of \citet{Collet07} to estimate that a typical 3D correction for subsample 3 stars might lower the sulphur abundances by about 0.05--0.15\,dex, bringing our results in better agreement with previous studies. Negative corrections for the [\ion{S}{i}] line are also predicted for metal-poor dwarf stars \citep{Caffau07}.

Further investigations could test these possibilities from both the observational and theoretical sides. For example, a thorough study analysing the forbidden line, the $\lambda923$\,nm triplet and the $\lambda1045$\,nm triplet in an extended sample of cool metal-poor giants could firmly quantify the systematic differences between the diagnostics, assuming that they are real. Tailored non-LTE calculations using the most up to date model of the sulphur atom and the radiation field from MARCS models would test the differences of the non-LTE aspects of the lines more consistently. More calculations on the line formation of the three diagnostics in 3D stellar atmospheres would allow us to better quantify how reducing the problem to one dimension affects the lines. 

   Systematic effects concerning the different methods of deriving stellar parameters also deserve a closer inspection. Note that the 65~K difference between our $T_\mathrm{eff}$ scale and the IQM scale makes the sulphur abundances derived in this work lower by about 0.02~dex when compared to the hypothetical case of no difference. 
   
\subsubsection{Comparison with theoretical models}\label{ch:disc_sfe_gce}

\begin{figure*}
   \sidecaption
   \includegraphics[width=12cm]{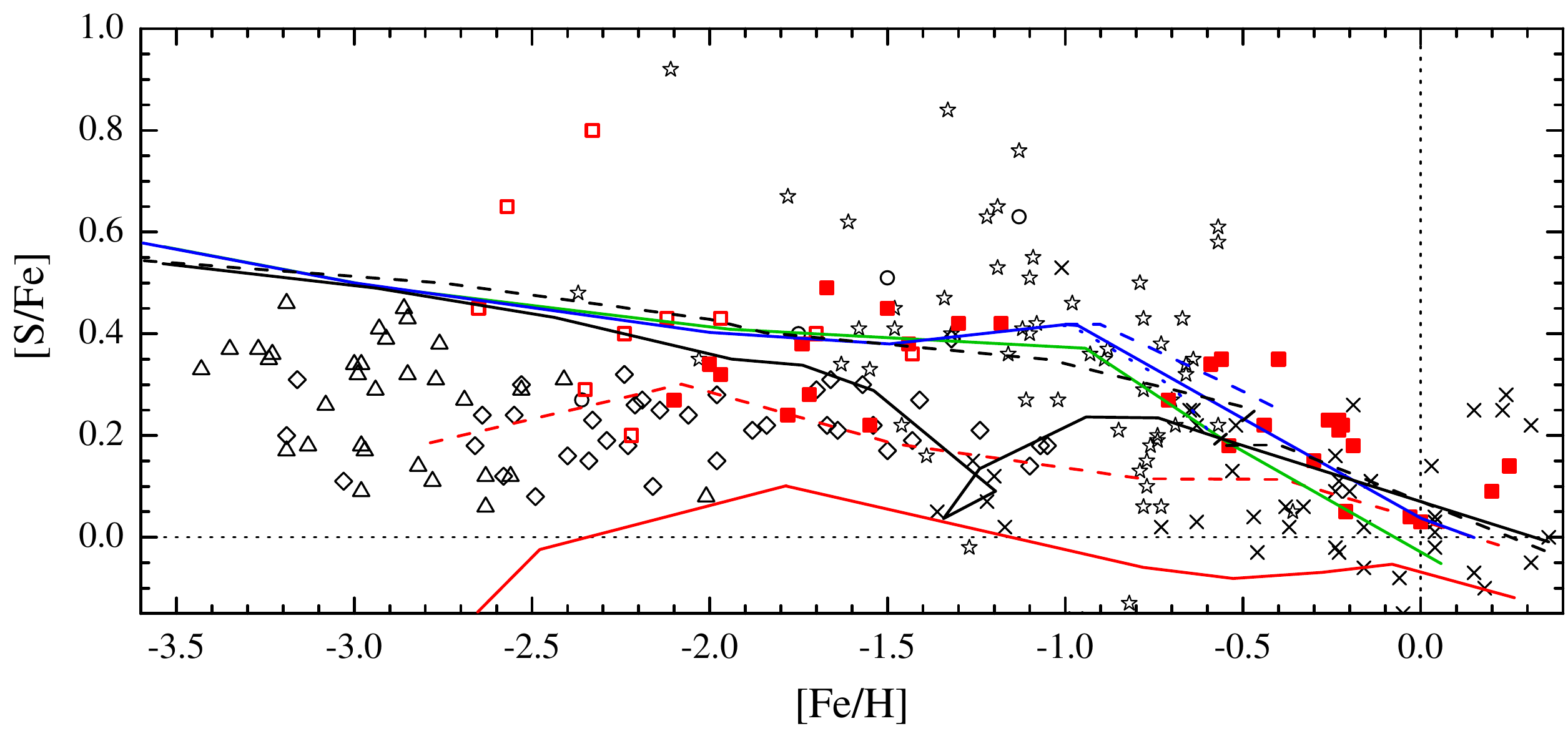}
      \caption{Comparison to theoretical models of Galactic chemical evolution. The symbols corresponding to previous studies are the same as in Fig.~\ref{fig:sfeprev}. Our results are plotted with solid and empty red squares. The red solid and dashed lines correspond to \citet{Timmes95} nominal model and one where the iron yields are reduced by a factor of two, respectively. The green line is the prediction of \citet{Kobayashi06} for the solar neighbourhood. The dotted, solid and dashed blue lines correspond to Halo, solar neighbourhood and Thick disk models of \citet{Kobayashi11}, respectively. The solid and dashed black lines correspond to the \citet{Brusadin13} double infall model with and without outflow, respectively.}
         \label{fig:sfe_gce}
   \end{figure*}

In Fig.~\ref{fig:sfe_gce} we compare our results to the predictions of some of the Galactic chemical evolution models that have been published in literature. In particular, we consider the models by \citet{Timmes95}, \citet{Kobayashi06}, \citet{Kobayashi11} and \citet{Brusadin13}.

\citet{Timmes95} consider a simple Galactic chemical evolution model where the Galaxy is formed by a collapse of a massive rotating gas cloud onto an exponential disk and a $1/r^2$ bulge with a gas infall e-folding timescale of 4~Gyr. They assume a \citet{Salpeter55} initial mass function, a quadratic \citet{Schmidt63} star formation rate, and instantaneous mixing of yields back into the interstellar medium. They adopt Type~II supernovae (SNe) yields from their companion paper \citep{Woosley95}, and yields from low-mass stars and Type~Ia SNe from previous works by other authors. According to this model at low metallicities sulphur is produced primarily in explosive oxygen burning in Type~II SNe. 

Even though some of the assumptions can now be considered outdated, the model manages to reproduce the observed abundance trends for quite a few chemical elements between hydrogen and zinc when, as \citet{Timmes95} themselves concluded, the iron yields calculated by \citet{Woosley95} are reduced by a factor of 2. Of note is the fact that the model predicts a fairly shallow decline of $\mathrm{[S/Fe]}$ starting from $\mathrm{[Fe/H]}\sim-2$ (red lines in Figure~\ref{fig:sfe_gce}) and a low enrichment of sulphur at all metallicities. The $\mathrm{[S/Fe]}$ trend is relatively flat over the metallicity range between $-1<\mathrm{[Fe/H]}<0$ because the additional sulphur produced in the more metal-rich Type~II SNe (from progenitors with total metal content above 10\% solar) is balanced by the iron produced in Type~Ia SNe. 

\citet{Kobayashi06} simulate the chemical evolution of the solar neighbourhood (using a one-zone model), Galactic Bulge (infall model), Thick disk (infall model) and Halo (closed-box model). They use their own nucleosynthesis yields for Type~II supernovae and hypernovae, and the yields from their single-degenerate Type~Ia supernovae models. Motivated by observational constraints (the observed $\text{[O/Fe]}$ versus $\mathrm{[Fe/H]}$), they introduce a ``metallicity effect'' to delay the onset of SNe~Ia in their simulations until $\mathrm{[Fe/H]}=-1.1$.

Their simulations predict a slowly changing trend of $\mathrm{[S/Fe]}$ below $\mathrm{[Fe/H]}\simeq-1$, in the solar neighbourhood going from $\mathrm{[S/Fe]}\simeq0.5$ at $\mathrm{[Fe/H]}\simeq-3$ to $\mathrm{[S/Fe]}\simeq0.37$ at $\mathrm{[Fe/H]}\simeq-1$ due to metal-poor and massive SNe producing relatively larger amounts of sulphur. Above $\mathrm{[Fe/H]}\simeq-1$ an almost linear decrease of $\mathrm{[S/Fe]}$ is predicted (green solid line).

\citet{Kobayashi11} use the same models as \citet{Kobayashi06} but with some improvements, e.g., they use updated Type~II SNe yields, and their chemical evolution model includes yields from stellar winds and asymptotic giant branch stars \citep{Karakas10}. According to this and the \citet{Kobayashi06} model at low metallicities sulphur is chiefly produced in Type~II SNe with a minor contribution from hypernovae.

Their results predict sulphur trends (blue lines) that are in a very good agreement with their previous results for $\mathrm{[Fe/H]}<-1$, while a larger $\mathrm{[S/Fe]}$ is predicted at higher metallicities. The predicted differences between the solar neighbourhood (solid line), Thick disk (dashed line) and Galactic Halo (dotted line) are smaller than observed for other $\alpha$-elements with the Halo being at most 0.1~dex less enriched in sulphur than the Thick disk at around $\mathrm{[Fe/H]}=-0.5$. 

\citet{Brusadin13} adopt the two-infall model originally proposed by \citet{Chiappini97}, to which they add an outflow  (wind) from Halo stars. The two-infall model assumes that the Galaxy formed in two primary episodes of primordial gas accretion. During the first accretion episode the Halo and the Bulge are rapidly formed over a timescale of $\sim1$~Gyr. In the second episode of much slower accretion the Thin disk is formed ``from the inside out'', i.e., the accretion rate decreases with radius (the resulting timescale is $\sim7$~Gyr at solar galactocentric radius). The outflow is added to reproduce the observed metallicity distribution function of the Halo. \citet{Brusadin13} adopt Type~II SNe and hypernovae yields from \citet{Kobayashi06} and Type~Ia SNe yields from \citet{Iwamoto99}. In addition they include the yields from asymptotic giant branch stars \citep{Karakas10} and novae \citep{JoseHernanz98}.

The main difference between the models with (solid line) and without (dashed line) outflow is the knot at $\mathrm{[Fe/H]}\simeq-1.3$ predicted by the former, which is caused by a break in star formation not long after the first infall episode. This break coincides with the onset of the first Type~Ia SNe (at $\mathrm{[Fe/H]}\simeq-1.8$) and, as a result, there is a rapid decline in the $\text{[}\alpha\text{/Fe]}$-$\mathrm{[Fe/H]}$ plane to lower $\text{[}\alpha\text{/Fe]}$ values. Once the second infall episode starts, the overall metallicity decreases before star formation restarts and $\text{[}\alpha\text{/Fe]}$ again increases producing the knot.

When stars of all metallicities are considered, the results of this paper are best described by the predictions of the \cite{Kobayashi11} models, which better describe the results for $\mathrm{[Fe/H]}>-1$ than the \citet{Kobayashi06} and \citet{Brusadin13} models which predict a similar sulphur enrichment at low metallicities. The \citet{Timmes95} models are at odds with our results. However, the model with reduced iron yields (red dashed line) fits the results of \citet{Nissen07} very well. Of the two \citet{Brusadin13} models, the one with no outflow fits our results slightly better at $\mathrm{[Fe/H]}>-1$ and around the knot, but we only have a few stars with this metallicity. 

In stars with $-1.5\lesssim\mathrm{[Fe/H]}\lesssim-1$ sulphur abundances from $\mathrm{[S/Fe]}\sim0$ to $\mathrm{[S/Fe]}\sim0.8$ have been determined in different studies in the literature. The explanation for finding low enrichment of sulphur might be the sampling of a low-$\mathrm{[}\alpha\mathrm{/Fe]}$ Halo population, evidence for which has been found both from observations of Halo dwarfs \citep{Nissen10} and theoretical $\Lambda$CDM simulations of the formation and evolution of the Galactic Halo \citep{Zolotov10}, according to which these low-$\mathrm{[}\alpha\mathrm{/Fe]}$ stars have formed in satellite galaxies with lower star-formation rates and later accreted by the Milky Way. Distinct Halo populations are not incorporated in the models discussed in this section.

The theoretical $\mathrm{[S/Fe]}$ plateaus are slightly higher than we find in this study \citep[except for the models of][]{Timmes95} but the agreement with theory is generally better than with the studies of \citet{Nissen07} and \citet{Spite11}. The important point is that, although detailed quantitative comparison between the models and observations seems premature due to the relatively large uncertainties in both, our results seem to support the notion that the Galactic chemical evolution of sulphur at low metallicities can largely be explained by production in regular Type~II SNe (with, perhaps, some contribution from hypernovae) without the need of invoking additional mechanisms.

\section{Conclusions}

In this study we have derived the sulphur abundances (or their upper limits) from spectrum synthesis of the [\ion{S}{i}] $\lambda1082.1$~nm line in a sample of 39 stars spanning the stellar parameter ranges $3900\lesssim T_\mathrm{eff}\lesssim5200$~K, $0.5\lesssim\log{g}\lesssim3.0$ and $-2.4\lesssim\mathrm{[Fe/H]}\lesssim0.2$ by employing one-dimensional local thermodynamic equilibrium MARCS model atmospheres.

The sulphur abundances show almost no dependence on the metallicity for $\mathrm{[Fe/H]}\lesssim-1$, which is typical for $\alpha$-elements. This means that the continued rise of $\mathrm{[S/Fe]}$ below $\mathrm{[Fe/H]}\simeq-1$ found in some previous studies is not supported by this study, in qualitative agreement with contemporary models of Galactic chemical evolution. However, the uncertainties are large enough that the metallicity range $-1.5\lesssim\mathrm{[Fe/H]}\lesssim-1$ should be studied further with a much larger sample of stars.

We manage to derive homogeneous stellar parameters for 29 of the stars (12 with $\mathrm{[Fe/H]}<-1$) -- we use photometric colour-effective temperature calibrations to derive $T_\mathrm{eff}$, physical gravities with Bayesian estimation to derive $\log{g}$, and spectrum synthesis to derive $\mathrm{[Fe/H]}$.

Our temperatures, derived from calibrations of \citet{GHB09}, are systematically hotter than the average literature temperatures by about 65~K. At the same time, no systematic effects with respect to literature are found for the surface gravities and metallicities derived here.

The forbidden sulphur line is a valuable diagnostic of sulphur abundances in giant and subgiant stars with $T_\mathrm{eff}\lesssim5200$~K and $\mathrm{[Fe/H]}\gtrsim-2.3$. The line is insensitive to the assumption of LTE and changes in $T_\mathrm{eff}$, but sensitive to changes in surface gravity and metallicity, which makes a homogeneous determination of the stellar parameters necessary. When results across multiple studies are compared, the differences in stellar parameters should be taken into account.

The uncertainties in the stellar parameters are found to dominate the total uncertainties in sulphur abundances. Therefore, the very high S/N values of most of our spectra can be considered a luxury that in future observing campaigns might be sacrificed in favour of observing a larger sample of stars. A $\mathrm{S/N}>100$ should be enough to derive precise sulphur abundances.

The sulphur abundances derived in this work from the forbidden sulphur line at $\lambda1082.1$~nm are on average higher than the abundances derived from the triplet at $\lambda923$~nm in some previous studies \citep{Nissen07,Spite11}. This hints at systematic differences between the stellar parameter scales or the two sulphur diagnostics. In the latter case, we suspect these differences to be due to non-LTE and 3D effects, but further investigation is required.

A worthwhile undertaking would be obtaining a homogeneous set of stellar parameters for the stars in which very high (and low) sulphur abundances have been derived in the past, and re-determining the sulphur abundances in their atmospheres from the forbidden sulphur line when possible.

\begin{acknowledgements}
N.~R. is a Royal Swedish Academy of Sciences Research Fellow supported by a grant from the Knut and Alice Wallenberg Foundation. Funds from Kungl. Fysiografiska S\"allskapet i Lund and support from the Swedish Research Council, VR are acknowledged. The authors are grateful to L.~Lindegren, S.~Feltzing and referee E.~Caffau for providing valuable feedback that improved the quality of this paper.
\end{acknowledgements}

\bibliographystyle{aa}
\bibliography{sulphur_aph}

\end{document}